\documentclass[preprint,epsfig,amsmath,superscriptaddress]{revtex4}

\usepackage[dvips]{graphicx}
\usepackage{latexsym}
\usepackage{natbib}

\begin{document}

\bibliographystyle{apsrev}

\title{Identifying influential spreaders in complex networks}

\author{Maksim Kitsak} \affiliation{Center for Polymer Studies and
  Physics Department, Boston University, Boston, Massachusetts 02215,
  USA} \affiliation{Cooperative Association for Internet Data Analysis
  (CAIDA), University of California-San Diego, La Jolla, California
  92093, USA}
  
\author{Lazaros K. Gallos} \affiliation{Levich Institute
  and Physics Department, City College of New York, New York, New York
  10031, USA}

\author{Shlomo~Havlin}
\affiliation{Minerva Center and
Department of Physics, Bar-Ilan
University, Ramat Gan, Israel}

\author{Fredrik~Liljeros}
\affiliation{Department of Sociology, Stockholm University, S-10691,
Stockholm, Sweden}

\author{Lev~Muchnik}
\affiliation{Information Operations and Management Sciences Department,
Stern School of Business, New York University, New York, New York 10012, USA}

\author{H. Eugene Stanley} \affiliation{Center for Polymer Studies and
  Physics Department, Boston University, Boston, Massachusetts 02215,
  USA}

\author{Hern\'an A. Makse}
\affiliation{Levich Institute and Physics
  Department, City College of New York, New York, New York 10031, USA}

\date{\today}

\maketitle

  \noindent {\bf Networks portray a multitude of interactions through
    which people meet, ideas are spread, and infectious diseases
    propagate within a society
    \cite{caldarelli,diseases,epidemiology,modelling,rogers}.
    Identifying the most efficient ``spreaders'' in a network is an
    important step to optimize the use of available resources and
    ensure the more efficient spread of information. Here we show
    that, in contrast to common belief,
    there are plausible circumstances where the best spreaders
    do not correspond to the best connected people
    or to the most central people (high betweenness
    centrality)~\cite{albert00,satorras01,cohen01,freeman,friedkin91}.
    Instead, we find: {\it (i)} The most efficient spreaders are those
    located within the core of the network as identified by the
    $k$-shell decomposition analysis \cite{kcore1,kcore2,carmi}. {\it
      (ii)} When multiple spreaders are considered simultaneously, the
    distance between them becomes the crucial parameter that
    determines the extent of the spreading. Furthermore, we find
    that infections persist in the high
    $k$-shells of the network,
    even in the case where recovered individuals do not develop immunity.
    Our analysis provides a
    plausible route for an optimal design of efficient dissemination
    strategies.}

Spreading is a ubiquitous process which describes many important
activities in society
\cite{diseases,epidemiology,modelling,rogers}. The knowledge of the
spreading pathways through the network of social interactions is
crucial for developing efficient methods to either hinder spreading in
the case of diseases, or accelerate spreading in the case of
information dissemination. Indeed, people are connected according to
the way they interact with each other in society and the large
heterogeneity of the resulting network greatly determines the
efficiency and speed of spreading. In the case of networks with a
broad degree distribution (number of links per node) \cite{albert00},
it is believed that the most connected people (hubs) are the key
players being responsible for the largest scale of the spreading
process \cite{satorras01,albert00,cohen01}.
Furthermore, in the context of social network theory, the importance
of a node for spreading is often associated with the betweenness
centrality, a measure of how many shortest paths cross through this node,
which is believed to determine who has more `interpersonal influence' on others \cite{freeman,friedkin91}.

Here we argue that the topology of the network organization plays an
important role such that there are plausible circumstances under which
the highly connected nodes or the highest betweenness nodes have
little effect in the range of a given spreading process.
For example, if a hub exists at the end of a branch at the periphery
of a network, it will have a minimal impact in the spreading process
through the core of the network, while a less connected person who is
strategically placed in the core of the network will have a
significant effect that leads to dissemination through a large
fraction of the population. In order to identify the core and the
periphery of the network we use the $k$-shell (also called $k$-core) decomposition  of the network
\cite{kcore1,kcore2,carmi,serrano}.  Examining
this quantity in a number of real networks allows us to identify the
best individual spreaders in the network when the spreading originates
in a single node. For the case of a spreading process originating in
many nodes simultaneously we show that we can further improve the
efficiency by considering spreading origins located at a determined
distance from each other.

We study real-world complex networks that represent archetypical
examples of social structures.  We investigate (i) the friendship
network between 3.4 million members of the {\it LiveJournal.com}
community \cite{lj}, (ii) the network of email contacts in the
Computer Science Department of the University College London (Zhou, S., private communication), (iii) the
contact network of inpatients (CNI) collected from hospitals in Sweden
\cite{hospital}, and (iv) the network of actors who have co-starred in
movies labeled by imdb.com as adult \cite{imdb} (see Supplementary
Information Section \ref{datasets} for details).

To study the spreading process we apply the
Susceptible-Infectious-Recovered (SIR) and
Susceptible-Infectious-Susceptible (SIS) models
\cite{diseases,epidemiology,hethcote00} on the above networks
(see Methods section). These models have been used to describe disease spreading
as well as information and rumor spreading in social processes where
an actor constantly needs to be reminded \cite{castellano09}.
We denote the probability that an infectious node will infect a susceptible
neighbor as $\beta$. In our study we use relatively small values for $\beta$,
so that the infected percentage of the population remains small.
In the case of large $\beta$ values, where spreading  can reach
a large fraction of the population, the role of individual nodes is no longer
important and spreading would cover almost all the network, independently of
where it originated from.

The location of a node in the network is obtained using the $k$-shell
decomposition analysis \cite{kcore1,kcore2,carmi}. This process
assigns an integer index or coreness, $k_S$, to each node representing
its location according to successive layers ($k$-shells) in the network.
The $k_S$ index is a quite robust measure and the nodes
ranking is not influenced significantly in the case of incomplete
information
(for details see SI-Fig.~\ref{ks_robustness} in SI-Section \ref{kshell}).
Small values of $k_S$ define the periphery of the
network and the innermost network core corresponds to large $k_S$ (see
Fig.~\ref{histograms}a and SI-Section
\ref{kshell}). Figures \ref{histograms}b-d
illustrate the fact that the size of the population infected in a
spreading process (shown in this example in the CNI network) is not
necessarily related to the degree of the node, $k$, where the
spreading have started. Spreading may be very different even when it
starts from hubs of similar degree as comparatively shown in
Figs.~\ref{histograms}b and c.
Instead, the location of the spreading origin given by its $k_S$ index
predicts more accurately the size of the infected population.  For
instance, Figs.~\ref{histograms}b and \ref{histograms}d show that
nodes in the same $k_S$ layer produce similar spreading areas even if
they have different $k$ (by definition, in a given layer there could
be many nodes with $k \ge k_S$).

The above example suggests that the position of the node relative to
the organization of the network determines its spreading influence
more than a local property of a node, like the degree $k$. To quantify the
influence of a given node $i$ in an SIR spreading process we study the
average size of the population $M_i$ infected in an epidemic
originating at node $i$ with a given $(k_S,k)$. The infected
population is averaged over all the origins with the same $(k_S,k)$
values:
\begin{equation}
M(k_S,k)=\sum_{i \in \Upsilon(k_S,k)} \frac{M_i}{N(k_S,k)},
\end{equation}
where $\Upsilon(k_{S},k)$ is the union of all $N(k_{S},k)$ nodes with
$(k_{S},k)$ values.

The analysis of $M(k_S,k)$ in the studied social networks reveals
three general results (see Fig.~\ref{incidence}): (a) For a fixed
degree, there is a wide spread of $M(k_S,k)$ values. In particular,
there are many hubs located in the periphery of the network (large
$k$, low $k_S$) that are poor spreaders.  (b) For a fixed $k_S$,
$M(k_S,k)$ is approximately independent of the degree of the
nodes. This result is revealed in the vertically layered structure of
$M(k_S,k)$ suggesting that infected nodes located in the same
$k$-shell produce similar epidemic outbreaks $M(k_S,k)$ independent of
the value of $k$ of the infection origin.  (c) The most efficient
spreaders are located in the inner-core of the network (large $k_S$
region) fairly independently of their degree.
These results indicate that the $k$-shell index of a node is a better
predictor of spreading influence. When an outbreak starts in the core
of the network (large $k_S$) there exist many pathways through which a
virus can infect the rest of the network; this result is valid
regardless of the node degree.
The existence of these pathways implies that during a typical epidemic
outbreak from a random origin, nodes located in high $k_S$ layers are
more likely to be infected and they will be infected earlier than other nodes
(see SI-Section~\ref{eandt}).  The neighborhood of these nodes
makes them more efficient in sustaining an infection at the early stages, 
allowing thus the epidemics to reach a critical mass that will allow it
to fully develop.
Similar results on the efficiency of high-$k_S$ nodes
are obtained from the analysis of $M(k_S,C_{B})$ in
Fig.~\ref{incidence}, where $C_B$ is the betweenness centrality of a
node in the network \cite{freeman,friedkin91}: the value of $C_B$ is
not a good predictor for spreading efficiency.

To quantify the importance of $k_S$ in spreading
we calculate the ``imprecision functions'' $\epsilon_{k_S}(p)$,
$\epsilon_k(p)$, and $\epsilon_{C_B}(p)$.
These functions estimate for each of the three indicators
$k_S$, $k$, and $C_{B}$ how close to the optimal spreading
is the average spreading of the $pN$ ($0<p<1$) chosen origins in each case, 
(see Methods and SI-Section \ref{imprecision}).  The strategy to predict the spreading efficiency
of a node based on $k_S$ is consistently more accurate than a method
based on $k$ in the studied $p$-range (Fig. \ref{scatter}a).  The
$C_{B}$-based strategy gives poor results compared to the other two
strategies.

Our finding is not specific to the social networks shown in
Fig. \ref{incidence}.  In SI-Section \ref{other} we analyze the
spreading efficiency in other networks not social in origin, like the
Internet at the router level \cite{dimes}, with similar
conclusions.  The key insight of our finding is that in the studied
networks a large number of hubs are located in the peripheral low
$k_S$ layers (Fig.~\ref{scatter}b shows the location of the 25
largest hubs in the CNI, see also SI-Section \ref{other}) and
therefore contribute poorly to spreading. The existence of hubs in the
periphery is a consequence of the rich topological structure of real
networks. In contrast, in a fully random network obtained by randomly
rewiring a real network preserving the degree of each node (such a
random network corresponds to the configuration model \cite{molloy},
see SI-Section \ref{rewiring}) all the hubs are placed in
the core of the network (see the red scatter plot in
Fig. \ref{scatter}c) and they contribute equally largely to spreading.
In such a randomized structure the same information is contained in
the $k$-shell as in the degree classification since there is a one to
one relation between both quantities which is approximately linear, $k_S
\propto k$ (Fig.~\ref{scatter}c and SI-Fig.~\ref{SIpowerlaw}).
Examples of real networks that are similar to a random structure
are the network of product space of economic goods \cite{hidalgo} and
the Internet at the AS level (analyzed in the SI-Section
\ref{other}).

Our study highlights the importance of the relative location of a {\it
single} spreading origin. Next, we address the question of the
extent of an epidemic that starts in {\it multiple} origins
simultaneously.  Figure \ref{scatter}d shows the extent of SIR
spreading in the CNI network when the outbreak simultaneously starts
from the $n$ nodes with the highest degree $k$ or the highest $k_S$
index. Even though the high $k_S$ nodes are the best single spreaders,
in the case of multiple spreading the nodes with highest degree are
more efficient than those with highest $k_S$.  This result is
attributed to the overlap of the infected areas of the different
spreaders: large $k_S$ nodes tend to be clustered close to each other,
while hubs can be more spread in the network and, in particular, they
need not be connected with each other. Clearly, the step-like features
in the plot of highest $k_S$ nodes (red solid curve in Fig. \ref{scatter}d)
suggest that the infected percentage remains constant as long as the
infected nodes belong in the same shell. Including just one
node from a different shell results in a significantly increased
spreading.  
This result suggests that a better spreading strategy using multiple
$n$ spreaders is to choose either the highest $k$ or $k_S$ nodes with
the requirement that no two of the $n$ spreaders are directly linked
to each other. This scheme then provides the largest infected area of
the network as shown in Fig. \ref{scatter}d.

Many contagious infections, including most sexually transmitted
infections \cite{yorke}, do not confer full immunity after infection
as assumed in the SIR model, and therefore are suitably described by
the SIS epidemic model, where an infectious node returns to the susceptible
state with probability $\lambda$. In an SIS epidemic the number of infectious
nodes eventually reaches a dynamic equilibrium ``endemic'' state where
as many infectious individuals become susceptible as susceptible nodes
become infectious \cite{hethcote00}.  
In contrast to SIR, in the initial state of our SIS simulations 20\%
of the network nodes are already infected. 
The spreading efficiency of a given node $i$ in SIS spreading is the
persistence, $\rho_{i}(t)$, defined as the probability that node $i$
is infected at time $t$ \cite{satorras01}. In an endemic SIS state,
$\rho_{i}(t\rightarrow\infty)$ becomes independent of $t$ (see
SI-Section \ref{persistance}). Previous studies have shown that the
largest persistence $\rho_{i}(t\rightarrow\infty)$ is found in the
network hubs which are re-infected frequently due to the large number
of neighbors \cite{satorras01,dezso02,satorras02}. However, we find
that this result holds only in randomized network structures. In the
real network topologies studied here, we find that viruses persist
mainly in high $k_S$ layers instead, irrespectively of the degree of
the nodes in the core.

In the case of random networks, it is found that viruses propagate to
the entire network above an epidemic threshold given by $\beta >
\beta^{\rm rand}_c \equiv \lambda \langle k\rangle/\langle
k^2\rangle$~\cite{cohen,satorras02}.  In real networks, such as the
CNI network, the threshold $\beta_c$ is different from $\beta^{\rm rand}_c$.
Furthermore, in real networks, we find that viruses can
survive locally even when $\beta<\beta_c$, but only within the high
$k_S$ layers of the network, while virus persistence in peripheral
$k_S$ layers is negligible (Fig.~\ref{sis}a-c). 
Since the $k$-shell structure depends on the network assortativity
the lower threshold is in agreement with the observation that
high positive assortativity \cite{newmanprl} may decrease the epidemic
threshold.

The importance of high $k_S$ nodes in SIS spreading 
is confirmed when we analyze the asymptotic probability that nodes of
given $(k_S,k)$ values will be infected. This probability is quantified by the persistence function
\begin{equation}
  \rho(k_S,k)\equiv\sum_{i\in\Upsilon(k_S,k)}\frac{\rho_i(t\rightarrow\infty)}{N(k_S,k)},
\end{equation}
as a function of $(k_S,k)$ at different $\beta$ values
(Fig.~\ref{sis}a and b).  
High $k_S$ layers in networks
might be closely related to the concept of a core group in Sexually
Transmitted Infections
research~\cite{yorke}. The core groups are defined as
subgroups in the general population characterized by high
partner turnover rate and extensive intergroup
interaction~\cite{yorke}.

Similar to the core group, the dense sub-network formed by nodes in
the innermost $k$-shells helps the virus to consistently survive locally
in the inner-core area and infect other nodes adjacent to the area.
These $k$-shells preserve the existence of a virus, in contrast to e.g.
isolated hubs in the periphery. Note that a virus cannot survive in the
degree-preserving randomized version of the CNI network, due to the
absence of high $k$-shells.

The importance of the inner-core nodes in spreading
is not influenced by the infection probability
values, $\beta$. In both models, SIS and SIR, we find that the
persistence $\rho$ or the average infected fraction $M$, respectively,
is systematically larger for nodes in inner $k$-shells compared to
nodes in outer shells, over the entire $\beta$ range that we studied
(Fig.~\ref{sis}c,d). Thus, the $k$-shell measure is a robust indicator
for the spreading efficiency of a node.

Finding the most accurate ranking of individual nodes for spreading in
a population can influence the success of dissemination
strategies. When spreading starts from a single node, the $k_S$ value
is enough for this ranking, while in the case of many simultaneous
origins, spreading is greatly enhanced when we additionally repel the
spreaders with large degree or $k_S$. In the case of infections that
do not confer immunity on recovered individuals, the core of the
network in the large $k_S$ layers forms a reservoir where infection can
survive locally.

\section{Methods}

\subsection{The $k$-shell decomposition}
Nodes are assigned to $k$-shells according to
their remaining degree, which is obtained by successive pruning of
nodes with degree smaller than the $k_S$ value of the current
layer. We start by removing all nodes with degree $k=1$.  After
removing all the nodes with $k=1$, some nodes may be left with one
link, so we continue pruning the system iteratively until there is no
node left with $k=1$ in the network. The removed nodes, along with the
corresponding links, form a $k$-shell with index $k_S=1$. In a similar
fashion, we iteratively remove the next $k$-shell, $k_S=2$, and
continue removing higher $k$-shells until all nodes are removed.  As a
result, each node is associated with a unique $k_S$ index, and the
network can be viewed as the union of all $k$-shells. The resulting
classification of a node can be very different than when the degree
$k$ is used.

\subsection{The spreading models}
To study the spreading process we apply the
Susceptible-Infectious-Recovered (SIR) and
Susceptible-Infectious-Susceptible (SIS) models. In the
SIR model, all nodes are initially in susceptible state (S) except for
one node in the infectious state (I). At each time step, the I nodes
attempt to infect their susceptible neighbors with probability $\beta$
and then enter the recovered state (R) where they become immunized and
cannot be infected again.  The SIS model aims to describe spreading
processes that do not confer immunity on recovered individuals:
infected individuals still try to infect their neighbors with probability
$\beta$ but they return to the susceptible state with probability
$\lambda$ (here we use $\lambda=0.8$) and can be reinfected at
subsequent time steps, while they remain infectious with probability
$1-\lambda$.

\subsection{The imprecision function}

The betweenness centrality, $C_B(i)$, of a node $i$ is defined as follows: Consider two
nodes $s$ and $t$ and the set $\sigma_{st}$ of all possible shortest paths
between these two nodes. If the subset of this set that contains the paths
that pass through the node $i$ is denoted by $\sigma_{st}(i)$, then the
betweenness centrality of this node is given by:
\begin{equation}
C_B(i) = \sum_{s\neq t} \frac{\sigma_{st}(i)}{\sigma_{st}},
\end{equation} 
where the sum runs over all nodes $s$ and $t$ in the network. 

The imprecision function $\epsilon(p)$ quantifies the difference in the average
spreading between the $pN$ nodes ($0<p<1$) with highest $k_S$, $k$, or
$C_B$ from the average spreading of the $pN$ most efficient spreaders
($N$ is the number of nodes in the network). Thus, it tests the
merit of using $k$-shell, $k$ and $C_{B}$ to identify the
most efficient spreaders. For a given $\beta$ value and a
given fraction of the system $p$ we first identify the set of the $Np$ most
efficient spreaders as measured by $M_i$ (we designate this set by
$\Upsilon_{\rm eff}$). Similarly, we identify the $Np$ individuals
with the highest $k$-shell index ($\Upsilon_{k_S}$).
We define the imprecision of $k$-shell identification as
$\epsilon_{k_{S}}(p) \equiv 1 - M_{k_S}/ M_{\rm eff}$,
where $M_{k_S}$ and $M_{\rm eff}$ are the average
infected percentages averaged over the $\Upsilon_{k_S}$ and
$\Upsilon_{\rm eff}$ groups of nodes
respectively. $\epsilon_{k}$ and $\epsilon_{C_{B}}$ are defined
similar to $\epsilon_{k_{S}}$.

\bigskip

\noindent{\bf Acknowledgements}

We thank NSF-SES, NSF-EF, ONR, Epiwork, and the Israel Science
Foundation for support. FL is supported by Riksbankens
Jubileumsfond. We thank L. Braunstein, J. Bruji\'c, kc claffy,
D. Krioukov, and C. Song for valuable discussions, and S. Zhou for
providing the email dataset.

\noindent{\bf Author contributions} 

All authors contributed equally to the work presented in this paper.

\noindent{\bf Additional information}

The authors declare no competing financial interests. Correspondence
and requests for materials should be addressed to H.A.M.

\clearpage

\begin{figure}
\includegraphics[width=17 cm,angle=0]{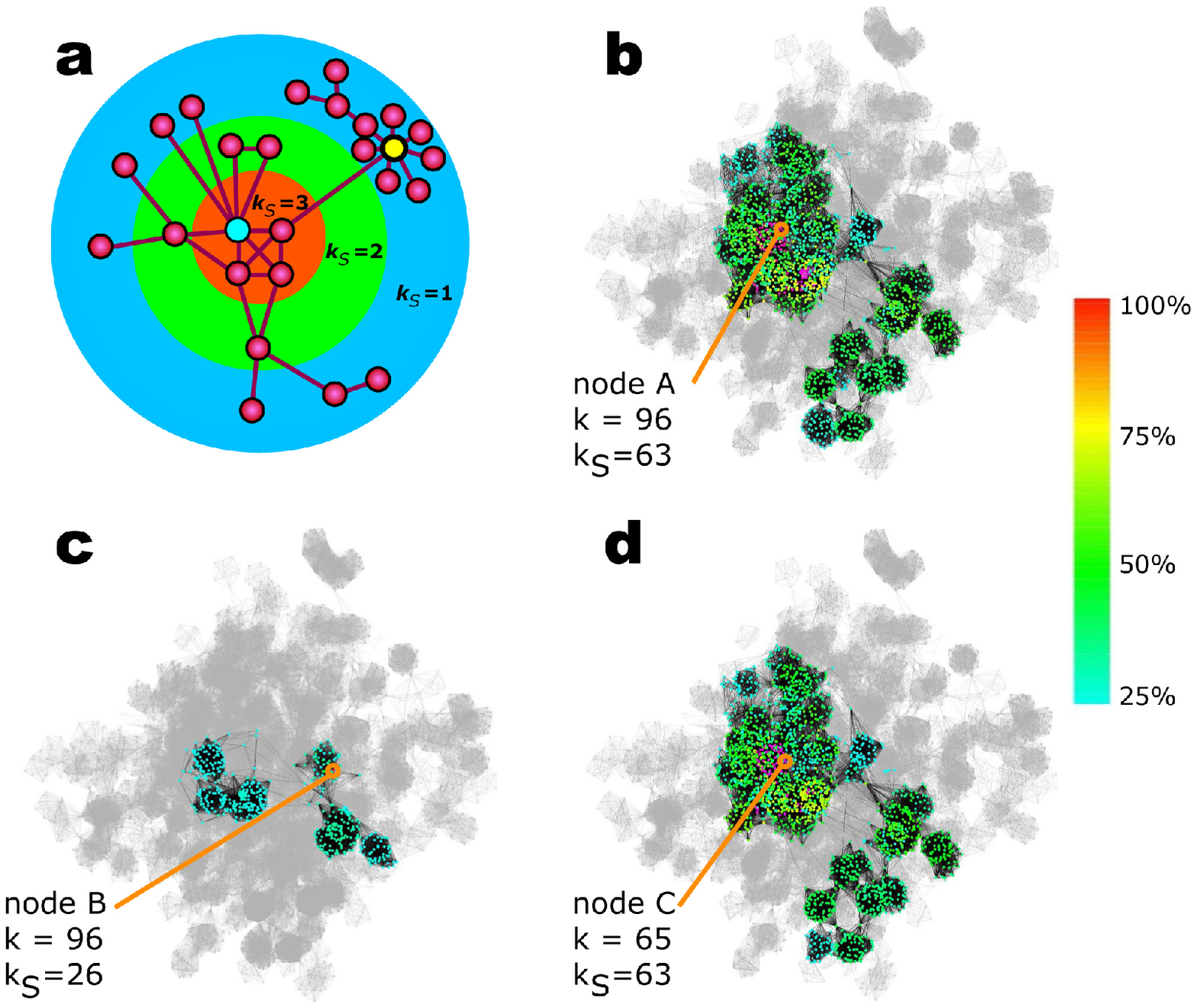}
\caption{}
\label{histograms}
\end{figure}

{\bf FIG \ref{histograms}. When the hubs may not be good spreaders.}
{\bf a,} A schematic representation of a network under the $k$-shell
decomposition. The two nodes of degree $k=8$ (blue and yellow nodes)
in this network are in different locations: one lies in the periphery,
($k_S=1$) while the other hub is in the innermost core of the network,
i.e. it has the largest $k_S$ ($k_S=3$).
{\bf b-d,} The extent of the efficiency of the spreading process cannot
be accurately predicted based on a measure of the immediate
neighborhood of the node, such as the degree $k$. For the contact
network of inpatients (CNI), we compare infections originating from
single nodes having the same degree $k=96$ (nodes A and B) or the same
index $k_S=63$ (nodes A and C), with infection probability
$\beta=0.035$.  In the corresponding plots, the colors indicate the
probability that a node will be infected when spreading starts in the
corresponding origin, as long as this probability is higher than 25\%.
The results are based on 10000 different realizations for each case.
In the first case, where origin A has $k_S=63$, spreading reaches a
much wider area more frequently, in contrast to origin B ($k_S=26$),
where the infection remains largely localized in the immediate
neighborhood of B. Spreading is very similar between origins A and C,
which have the same $k_S$ value, although the degree of C is much
smaller than A.  The importance of the network organization is also
highlighted when we randomly rewire the network (preserving the same
degree for all nodes).  In this case the standard picture is
recovered: the extent of spreading coincides and both hubs contribute
equally largely to spreading (see SI-Section \ref{rewiring}).

\begin{figure}
\includegraphics[height=6 cm]{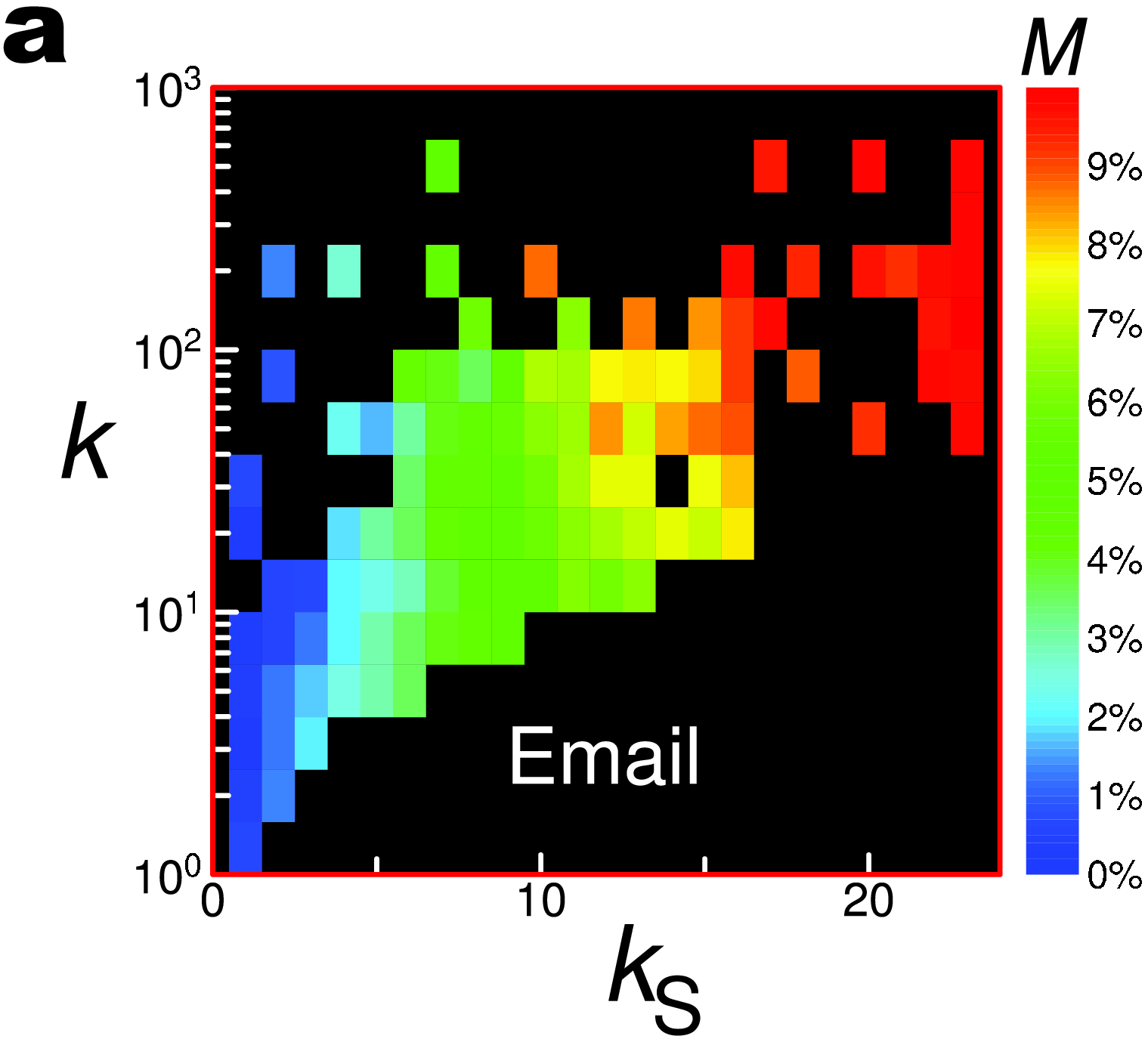}
\includegraphics[height=6 cm]{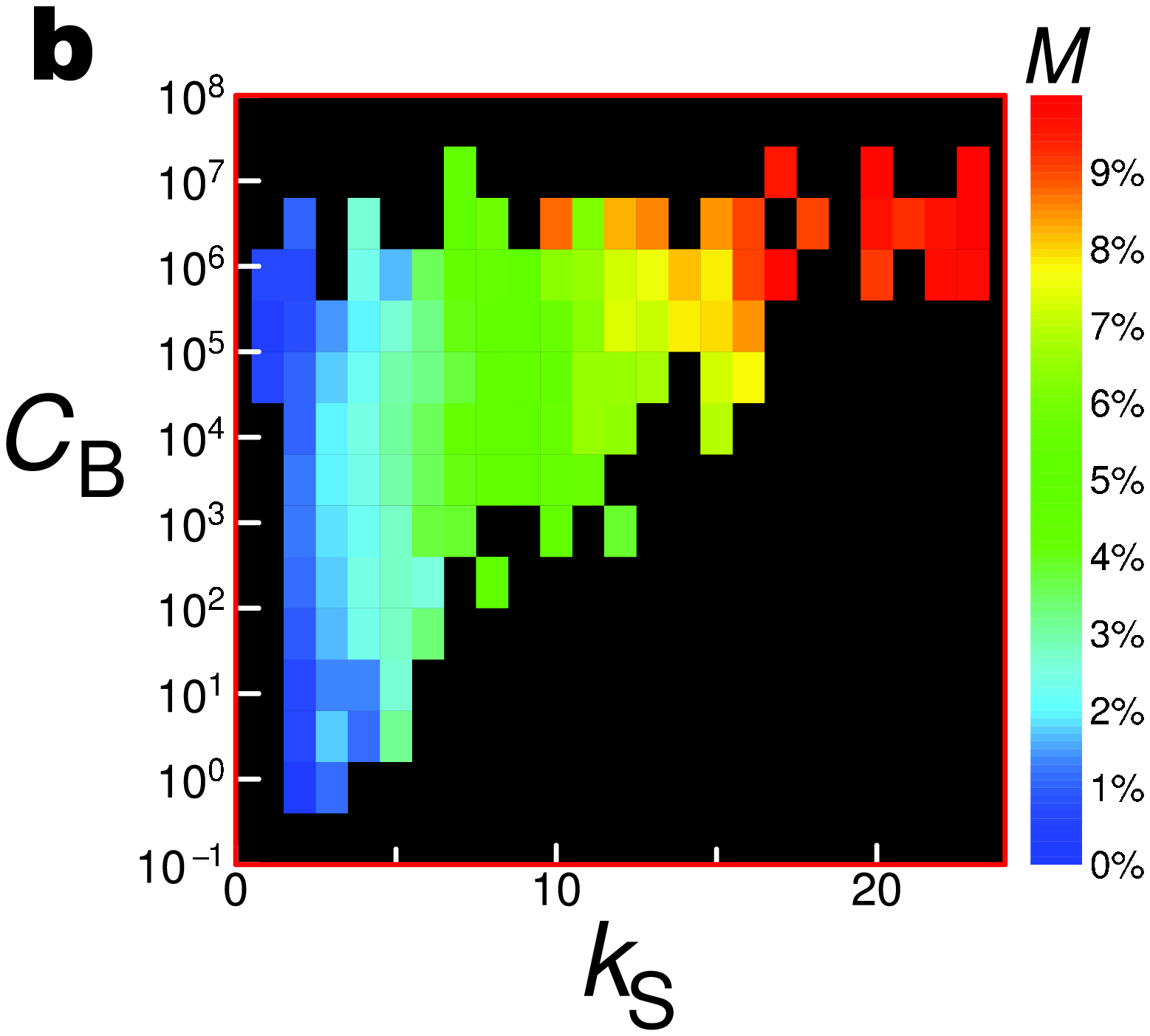}
\includegraphics[height=6 cm]{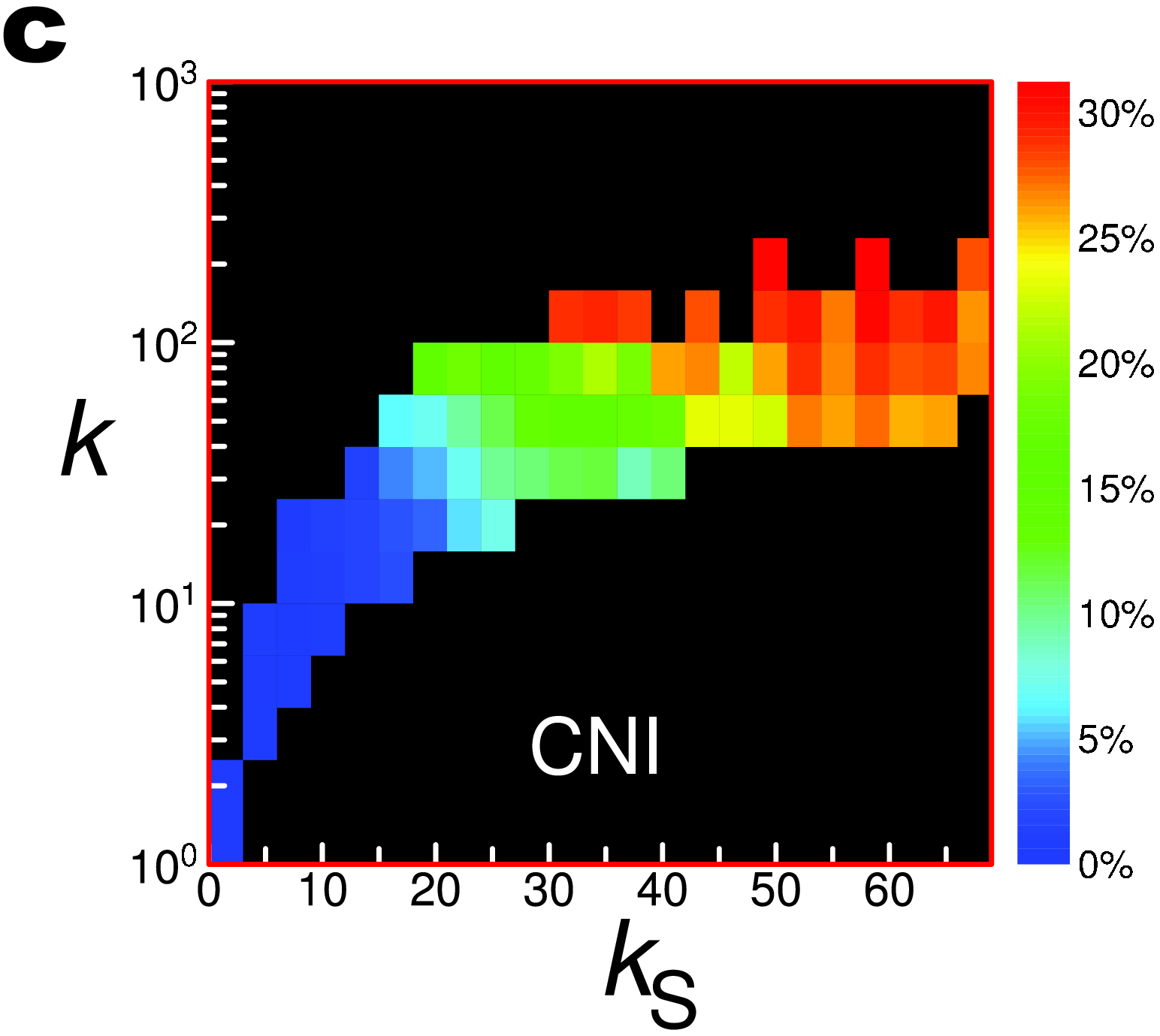}
\includegraphics[height=6 cm]{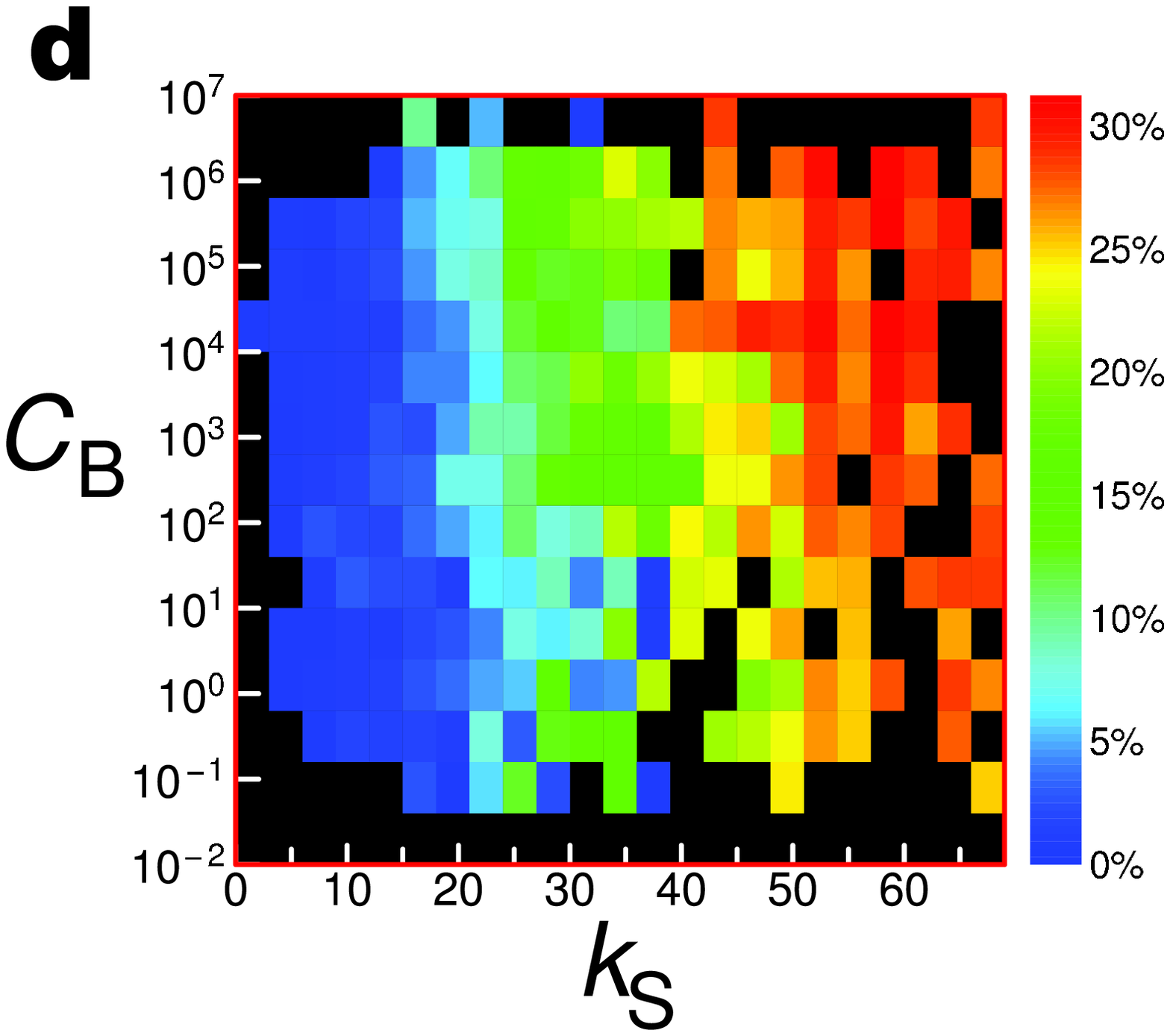}
\includegraphics[height=6 cm]{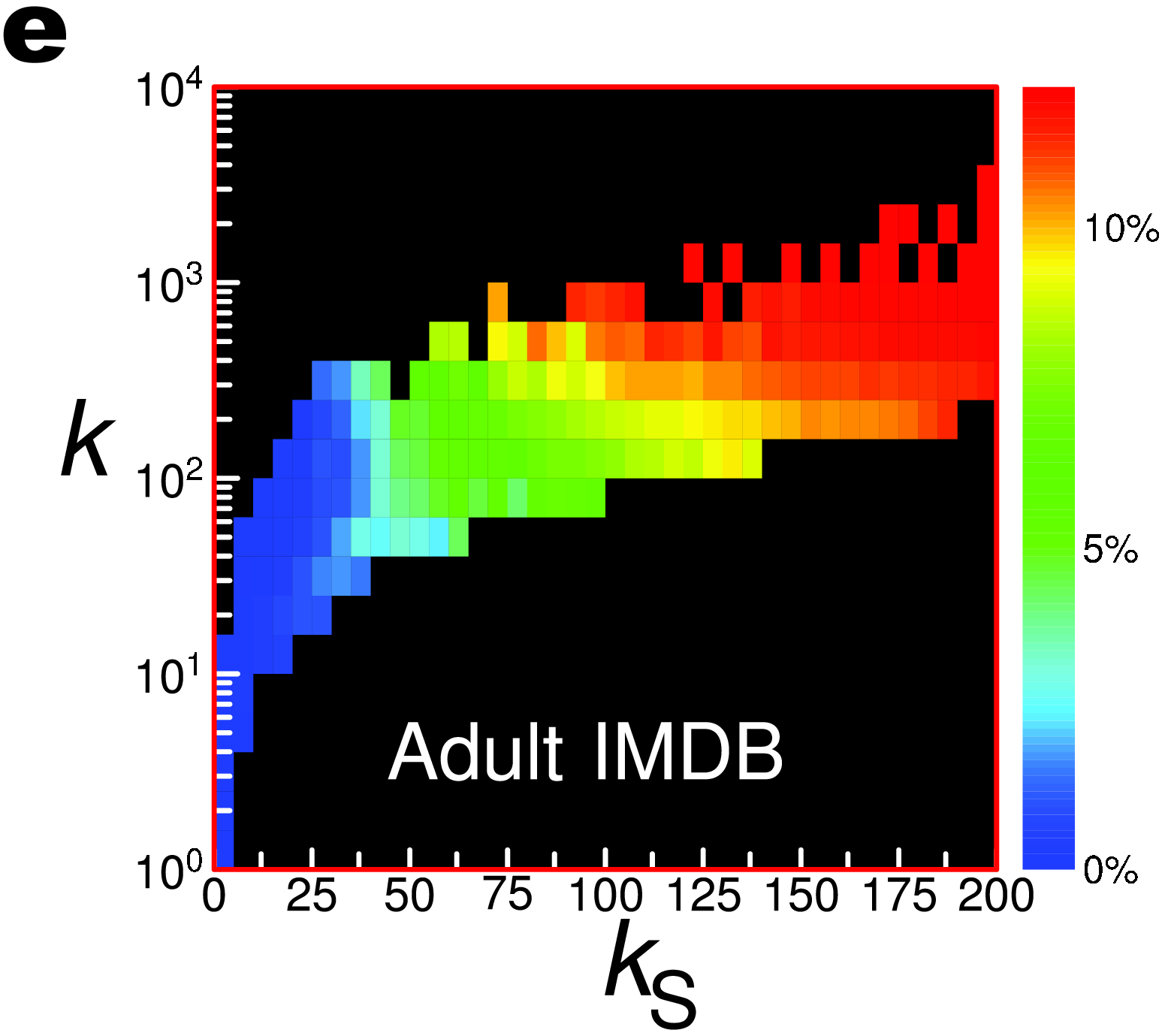}
\includegraphics[height=6 cm]{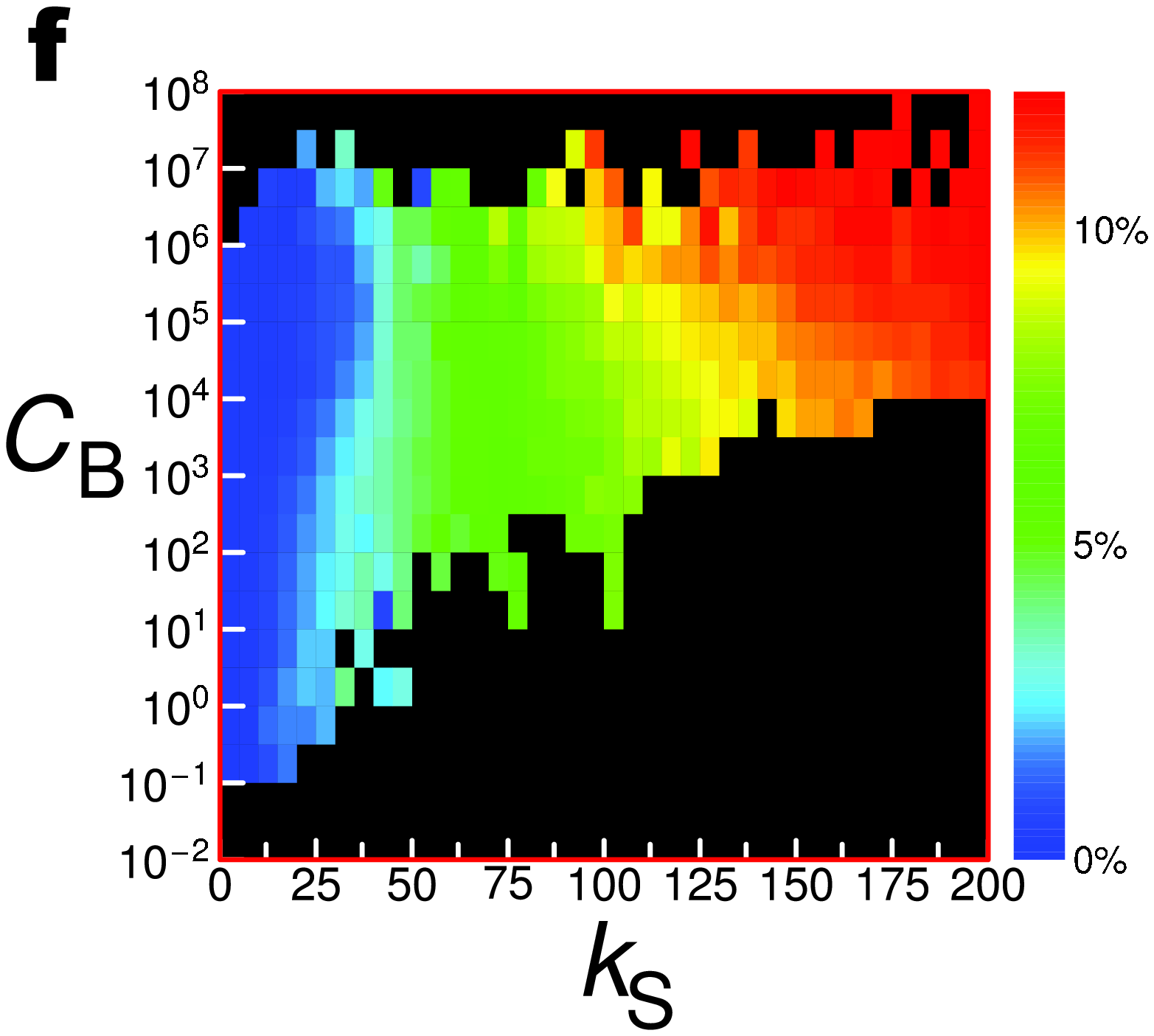}
\includegraphics[height=6 cm]{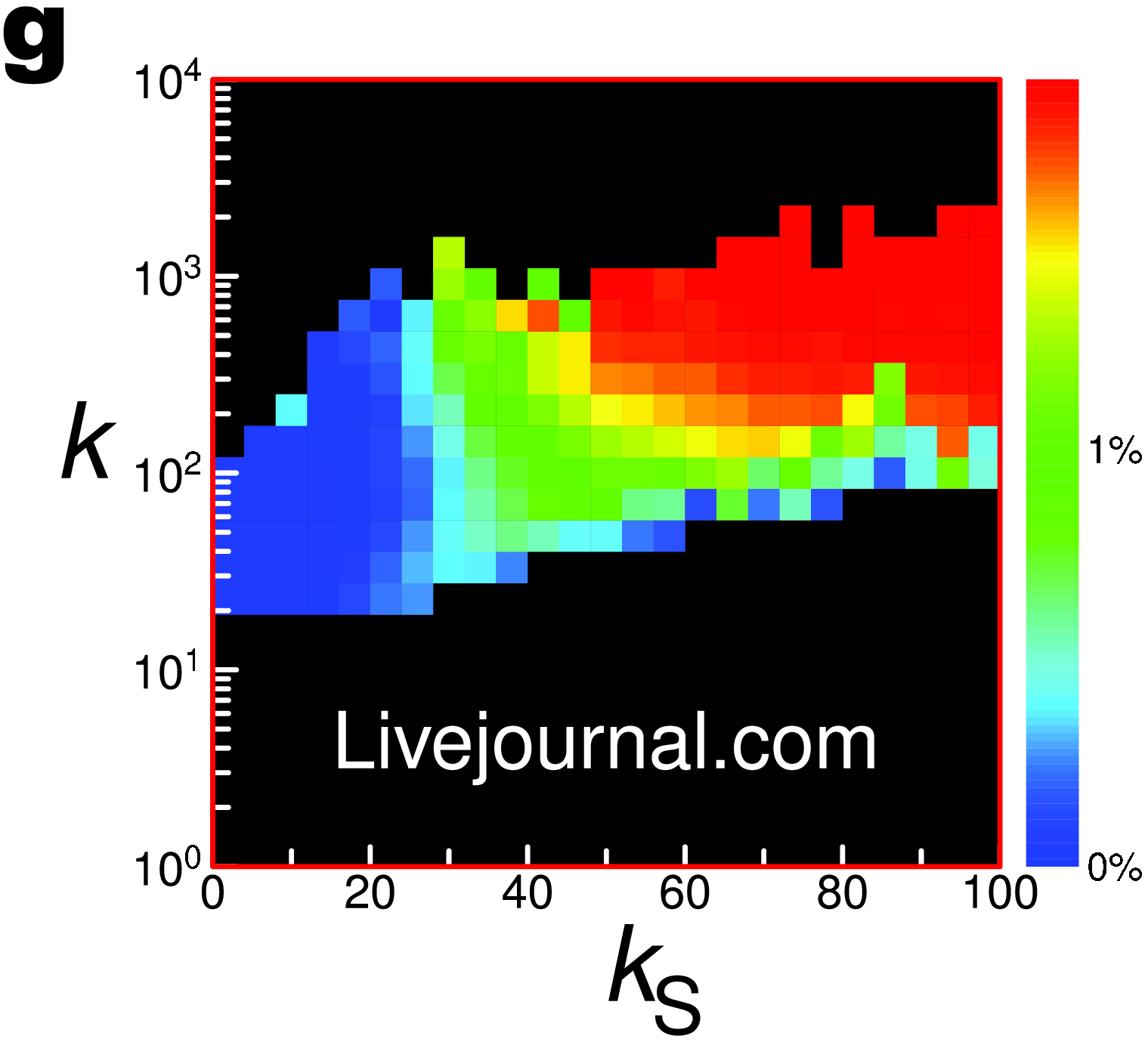}
\includegraphics[height=6 cm]{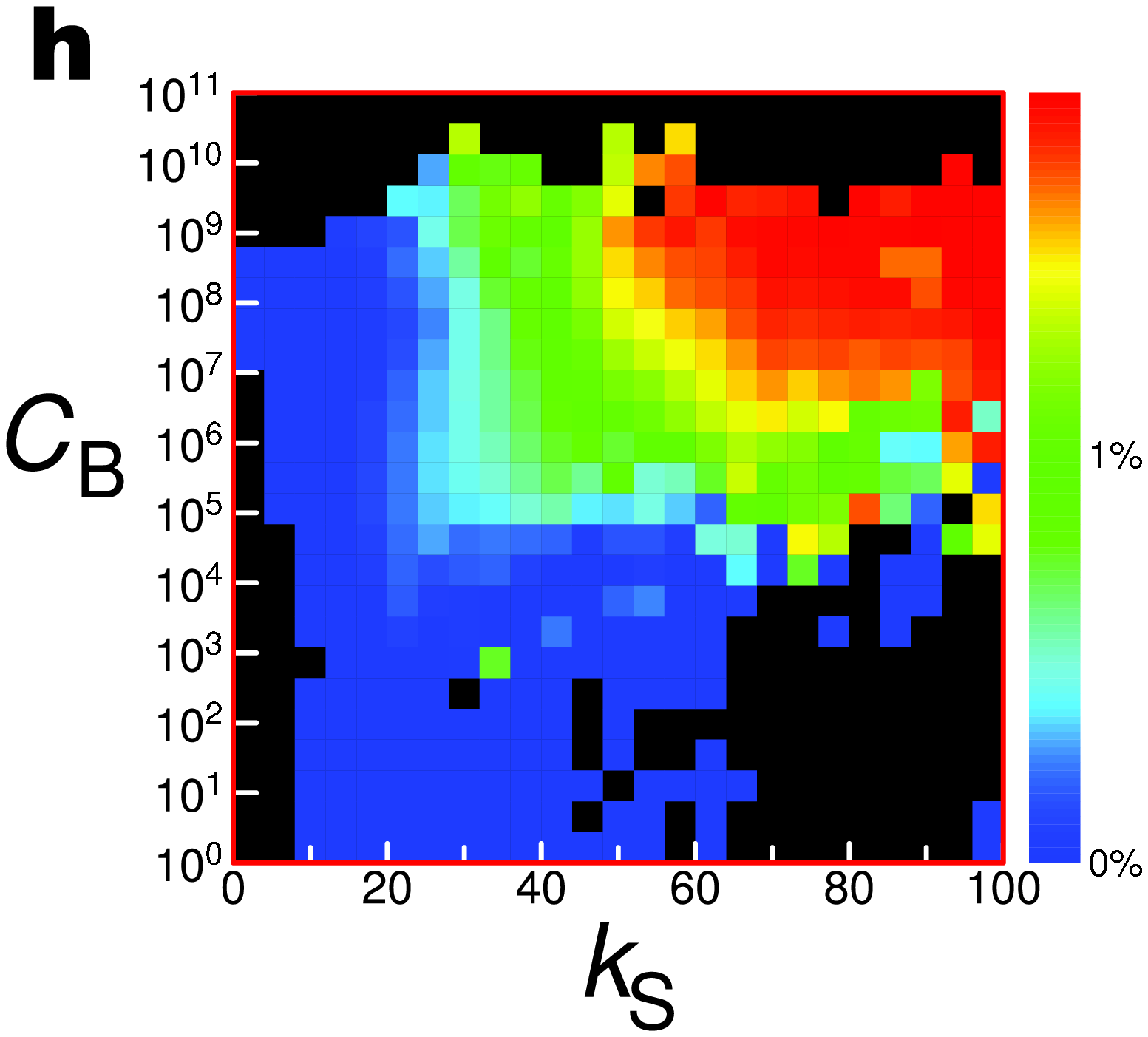}
\caption{}
\label{incidence}
\end{figure}

{\bf FIG \ref{incidence}. The $k$-shell index predicts the outcome of
  spreading more reliably than the degree $k$ or the betweenness
  centrality $C_B$.}  The networks used are (top to bottom): email
contacts ($\beta=8\%$), CNI network ($\beta=4\%$), the actors network
($\beta=1\%$), and the Livejournal.com friendship network
($\beta=1.5\%$). {\bf a, c, e, g} Average infected size of the
population $M(k_S,k)$ when spreading originates in nodes with
($k_S,k$). {\bf b, d, f, h} The infected size $M(k_S, C_B)$ when
spreading originates in nodes of a given combination of $k_S$ and
$C_B$.  In both cases, spreading is larger for nodes of higher $k_S$,
while nodes of a given $k$ or $C_B$ value can result in either small
or large spreading, depending on the value of $k_S$.  (There is an
exception at large $k_S$ and small $k$ of the livejournal database,
which is due to artificial closed groups of virtual characters that
connect with each other for the purpose of online gaming and do not
correspond to regular users, as the rest of the database.)

\begin{figure}
\center
\includegraphics[width=8cm,height=6cm,angle=0]{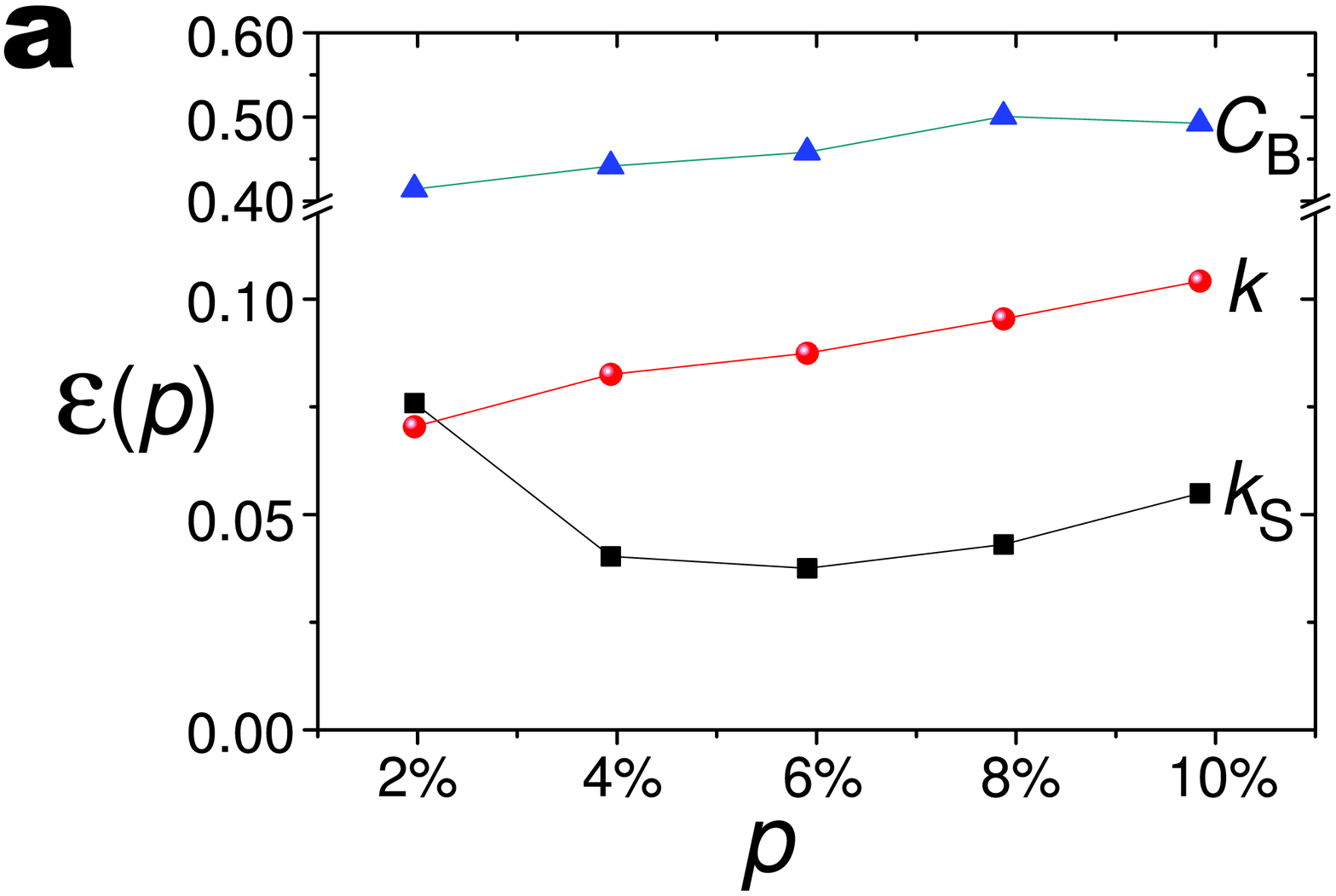}
\includegraphics[width=7cm,angle=0]{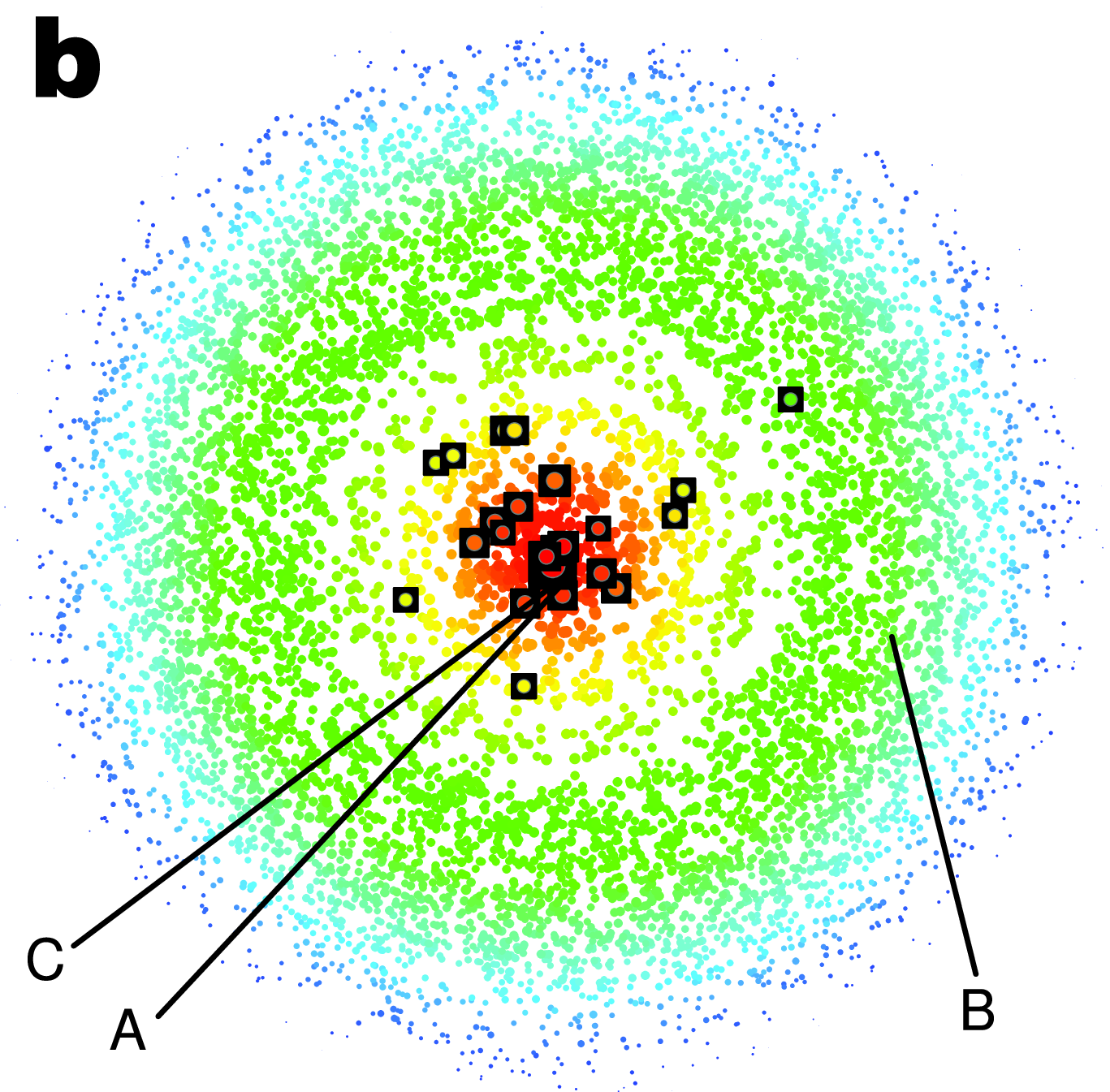}
\includegraphics[width=8cm,height=6cm,angle=0]{fig3c.eps}
\includegraphics[width=8cm,angle=0]{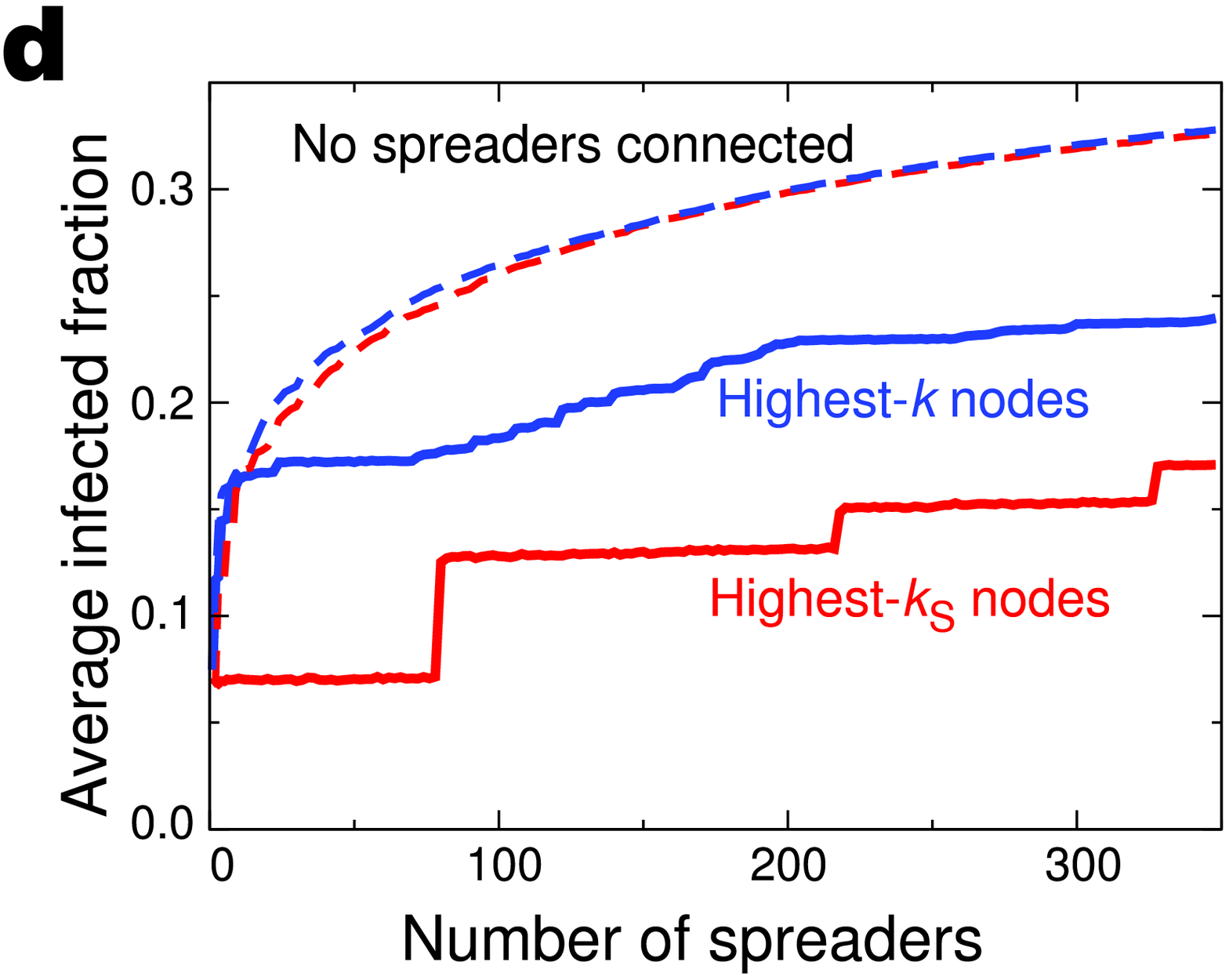}
\caption{}
 \label{scatter}
\end{figure}

{\bf FIG \ref{scatter}. $k$-shell structure of the CNI network.}  {\bf
  a,} The imprecision functions $\epsilon_{k_S}(p)$,
$\epsilon_{k}(p)$, and $\epsilon_{C_B}(p)$, for $\beta=4\%$.  Even
though both $k$-shell and $k$ identification strategies yield
comparable results for $p=2\%$, the $k$-shell strategy is consistently
more accurate for $2\% < p < 10\%$ with $\epsilon_{k_{S}}$
approximately twice lower than $\epsilon_{k}$.  The $C_{B}$
identification of the most efficient spreaders is the least accurate,
with $\epsilon_{C_{B}}$ exceeding $40\%$.  {\bf b,} We visualize the
CNI network as a set of concentric circles of nodes representing
inpatients, each circle corresponding to a particular $k$-shell.
The $k_S$ indices of a given layer increase as one moves
from the periphery to the center of the network
\cite{lanetvi,alvarezhamelin06}. Node size is proportional to the
logarithm of the degree of the node.  We highlight the 25 inpatients
with the largest degree values. Note that inpatients with high $k$
values are not concentrated at the ``center'' of the network but
instead are scattered throughout different $k$-shells.  We
highlight the position of the three nodes A, B, and C, of the origins
that were used in the example of Fig.~\ref{histograms}.  {\bf c,}
Scatter plot of the node degree $k$ as a function of $k_S$ for all the
nodes in the CNI network (black symbols) and the degree-preserving
randomized version of the same network (red symbols). Note that there
are many inpatients with large $k$ and low $k_S$ values in the
original network while in the randomized email network all the hubs
are located in the inner core of the network. We also show the
position of the three origins used in Fig.~\ref{histograms}.  {\bf d,}
When spreading starts from multiple origins, the set of nodes with
highest degree (blue continuous line) can spread significantly more
than the set of highest-$k_S$ nodes (red continuous line), because in
the latter case most of these nodes are connected to each other.  If
we only consider in this set nodes that are not directly linked, then
both the sets of highest $k$ or $k_S$ nodes yield a similar result
(dashed lines), where spreading is significantly enhanced. Results are
shown for $\beta=3\%$ in the CNI.

\begin{figure}
\center
\includegraphics[width=8cm,angle=0]{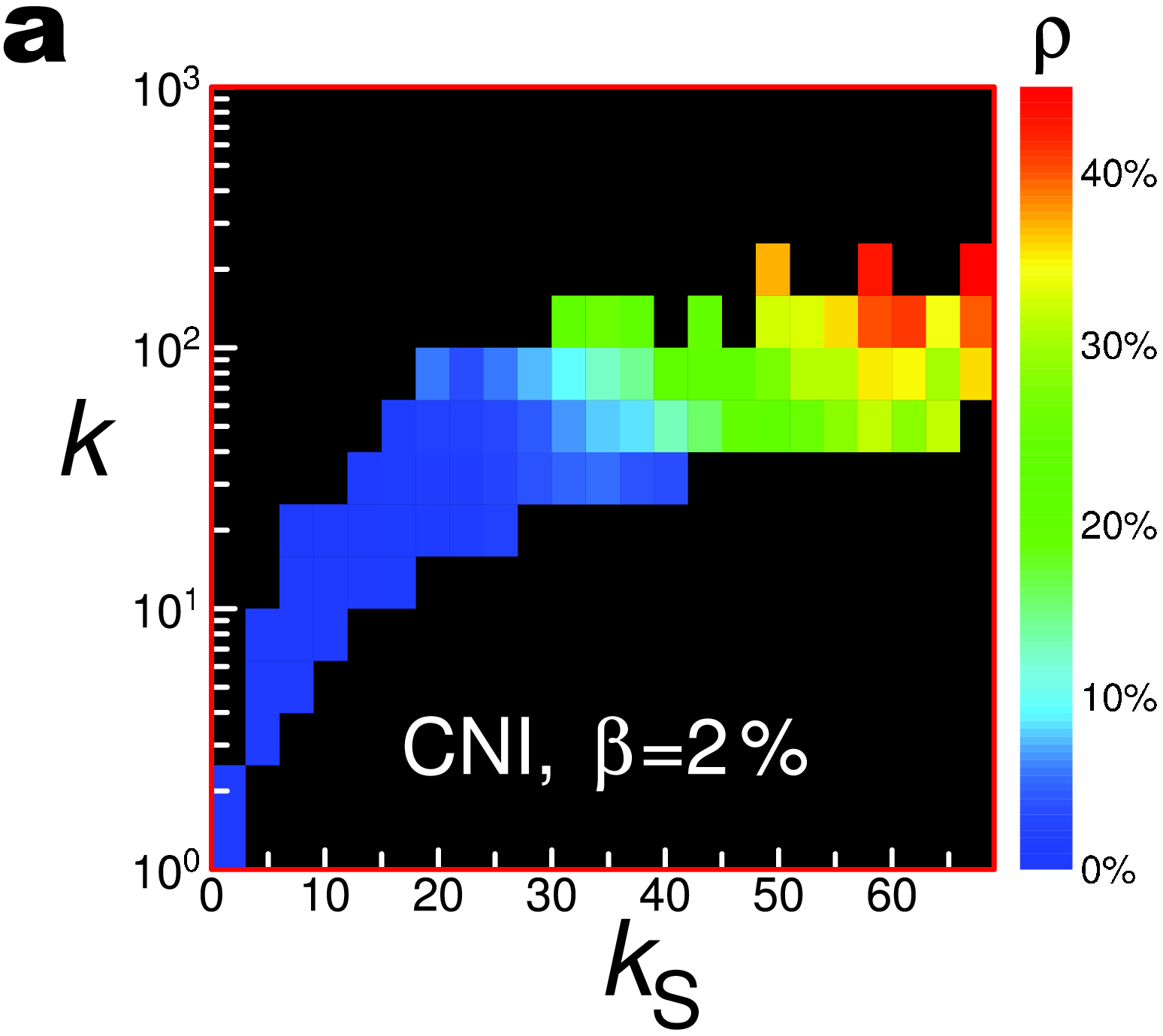}
\includegraphics[width=8cm,angle=0]{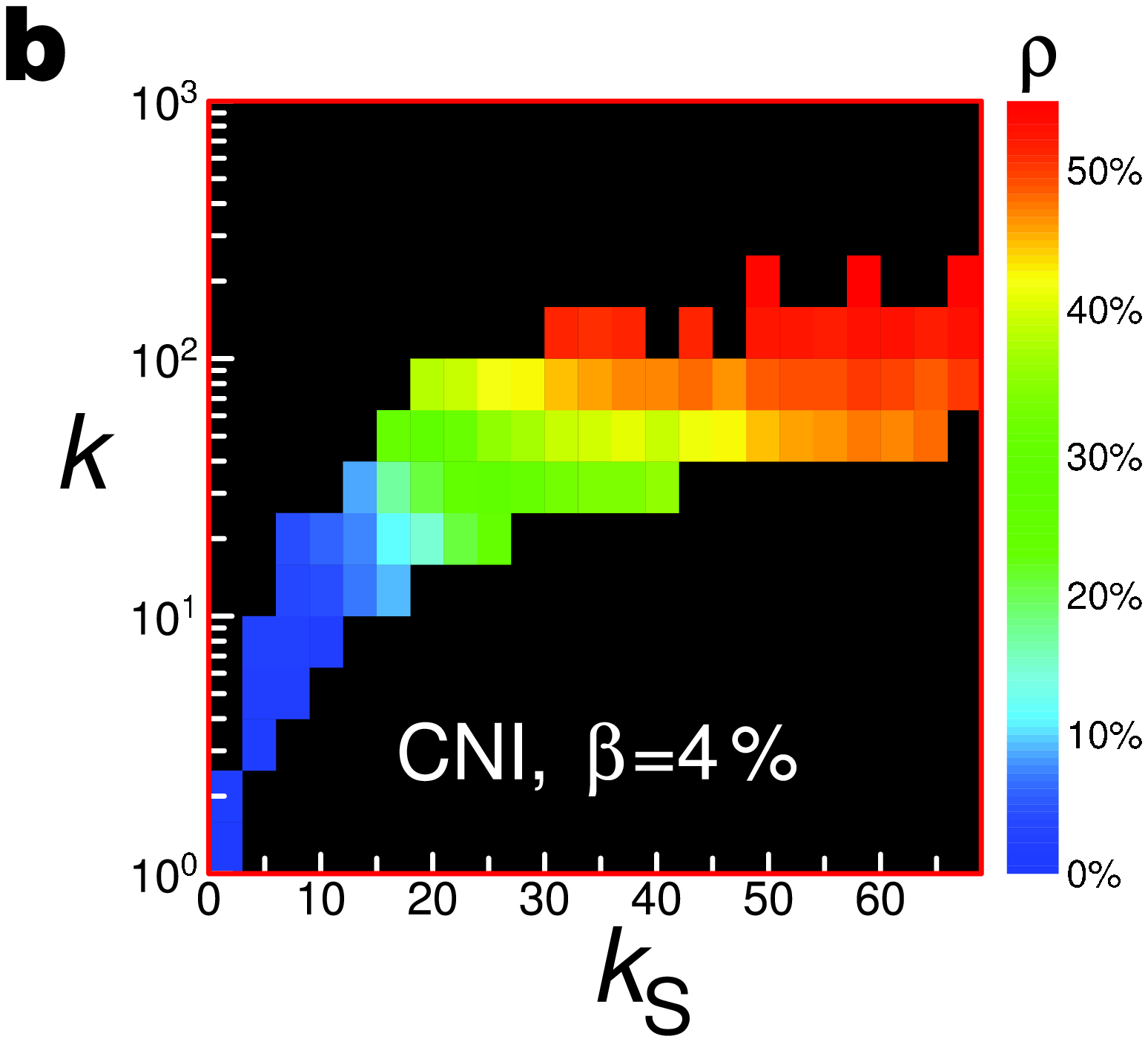}

\vspace{1cm}

\includegraphics[width=8cm]{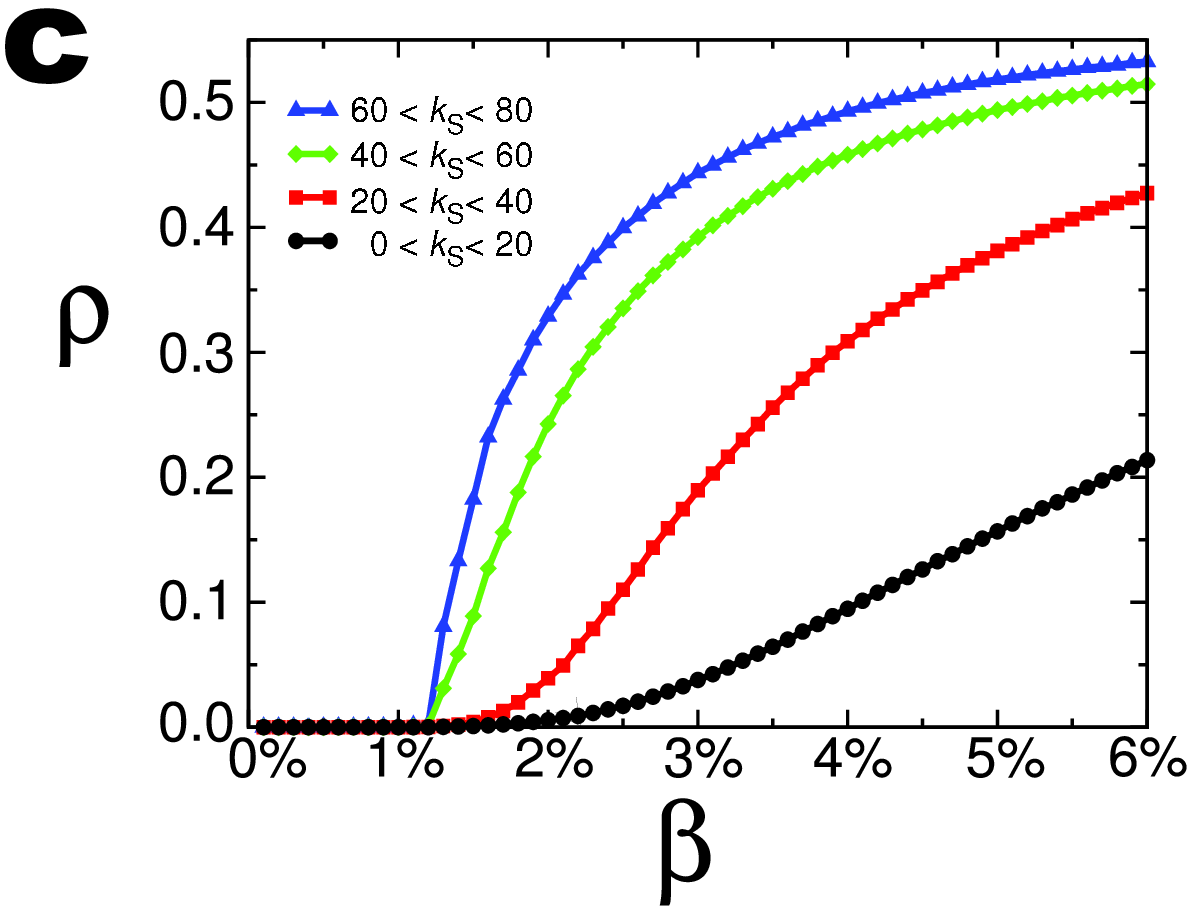}
\includegraphics[width=8cm]{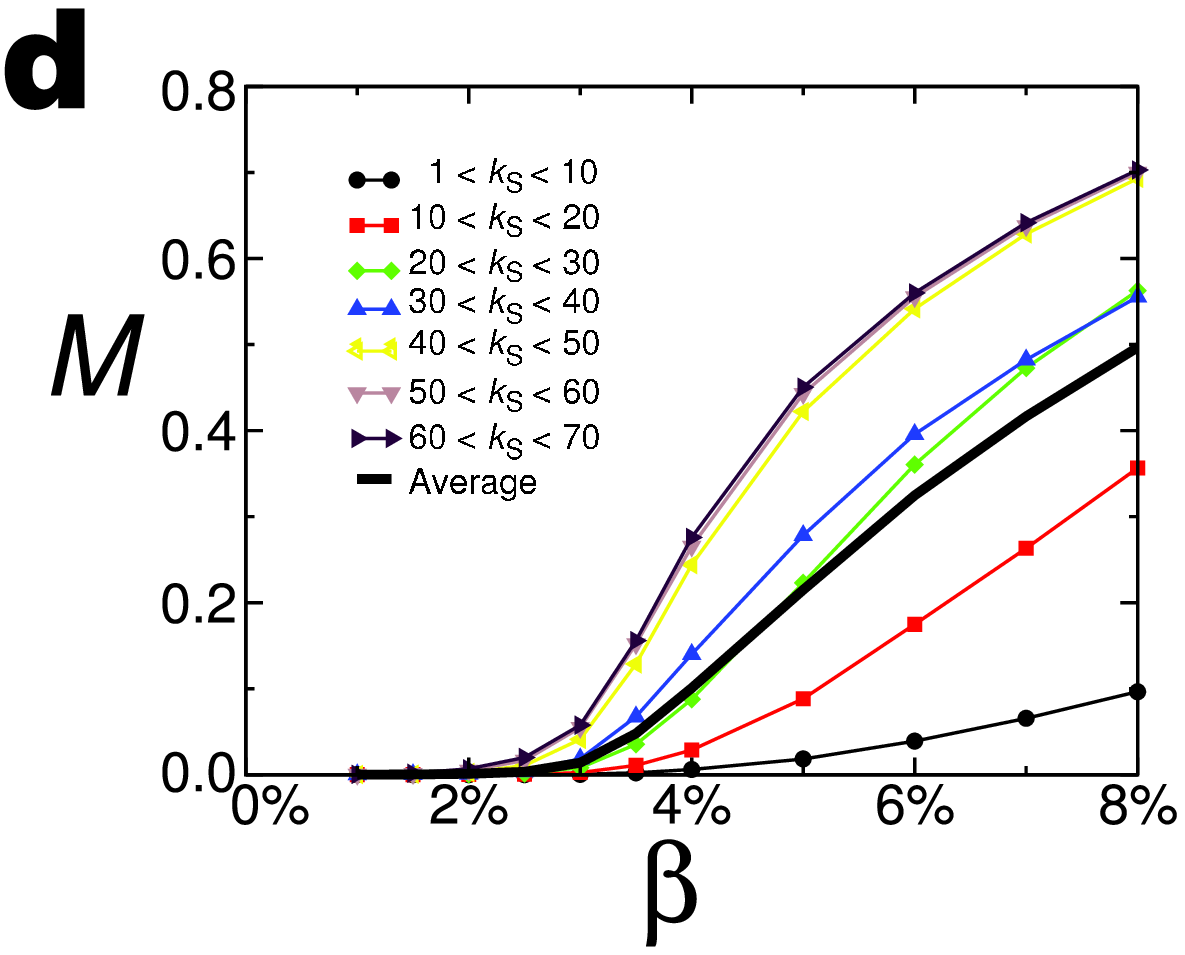}
\caption{}
 \label{sis}
\end{figure}

{\bf FIG \ref{sis}. SIS spreading in the CNI network and $\beta$
  dependence for SIS and SIR.} {\bf a, b,} Virus persistence
$\rho(k_{S},k)$ as a function of $k$ and $k_{S}$ values of inpatients
in the CNI network for, $\beta = 2\%$, and $\beta = 4\%$,
respectively, where 20\% of the individuals are initially infected.
The infection survives mainly in nodes with large $k_S$ values.  {\bf
  c,} We form four groups of nodes of the CNI network based on their
$k$-shell values. For all values of $\beta$, virus persistence is
consistently higher in the inner $k$-shells.
  {\bf d,} Influence of the infection probability $\beta$ on the spreading
  efficiency of nodes, grouped according to their $k$-shell values, for SIR
  spreading. The solid black line refers to the average infected percentage
  over all network nodes. Nodes in higher $k$-shells are consistently the most efficient,
  independently of the $\beta$ value.

\clearpage

\clearpage

\clearpage

\clearpage

\setcounter{section}{0}

\clearpage
\centerline{ \bf \Large Identifying influential spreaders in complex networks}
\vspace{.3cm}
\centerline{ \bf SUPPLEMENTARY INFORMATION}

\setcounter{page}{1}

\section{Datasets}
\label{datasets}

In this study we have mainly focused on social networks, but our results
can be extended to networks from practically any discipline. The
datasets that were used in the paper and in this Supplementary
Information are the following:

a) {\it Contact Network of Inpatients}. We use records from Swedish
hospitals \cite{hospital} and establish a link between two inpatients if they have both been
hospitalized in the same quarters. We restrict the recording period
to one week. All the data have been handled in a de-indentified form.
There are 8622 inpatients in the largest component, with an average
degree of around 35.1.

b) {\it IMDB actors in adult films}. We have created a network of
connections between actors who have co-starred in films, whose genre has
been labeled by the Internet Movie Database \cite{imdb} as `adult'.
This network is a largely isolated sub-set of the original actor
collaboration network.  Additionally, all these films have been produced
during the last few decades, rendering the network more focused in time.
The largest component comprises 47719 actors/actresses in 39397
films. The average degree of the network is 46.0.

c) {\it Email Contact Network}. The network of email contacts is based
on email messages sent and received at the Computer Sciences
Department of University College London. The data have been collected
in the time window between December 2006 and May 2007. Nodes in the
network represent email accounts.  We connect two email accounts with
an undirected link in the case where at least two emails have been
exchanged between the accounts (at least one email in each
direction). There are 12701 nodes with an average degree of 3.2.

d) {\it LiveJournal.com}. The network of friends in the LiveJournal
community, as recorded in a 2008 snapshot. We only consider reciprocal
links, i.e. when two members are in each other's list of friends.
There are 3453394 nodes in the largest component, and the average
degree is 12.4.

e) {\it Cond-mat collaboration network}. This is the network of
collaborations between scientists that have posted reprints in the
`cond-mat' e-print archive, between 1995 and 2005.
The nodes of the network represent the authors, who are connected if
they have co-authored at least one paper. The cond-mat collaboration
dataset consists of 17628 authors with average degree 6.0

\begin{table}
\begin{tabular}
 {|c|c|c|c|c|c|c|c|} \hline Network Name & $N$ & $N_E$ & $<k>$ &
 $<k^{2}>$ & $\beta^{\rm rand}_c$ & $\beta$ & $k_{S_{max}}$ \\
 \hline Contact Network of Inpatients & 8622 & 151649 & 35.1 & 1633 & 1.7\% & 4\%   & 66\\
 \hline Actor Network & 47719 & 1028537 & 46.0 & 17483 & 0.21\% & 1\%   & 199\\
 \hline Email Contacts & 12701 & 20417 & 3.2 & 351.1 & 0.73\% & 8\% & 23\\
 \hline Live Journal & 3453394 & 21378154 & 12.38 & 892.45 & 1.1\% & 1.5\% &  100 \\
 \hline Cond-mat Collaboration Network & 17628 & 52884 & 7.0 & 109.4 & 5.1\% & 10\% &  22 \\
 \hline RL Internet & 493312 & 808844 & 3.3 & 71.9 & 4.6\% & 6\% &  36\\
 \hline AS Internet & 20556 & 62920 & 6.1 & 2111.2 & 0.23\% & n/a & 41\\
 \hline Product Space & 765 & 40164 & 104.8 & 16931 & 0.50\% & n/a &  100 \\
 \hline
\end{tabular}
\caption{Properties of the real-world networks studied in this
  work. Here $N$ is the number of nodes, $N_E$ is the number of edges,
  $<k>$ is the average degree in the network, $<k^{2}>$ is the average
  squared degree in the network, $\beta^{\rm rand}_c$ is the epidemic threshold for a corresponding random
  network ($\beta^{\rm rand}_c \approx \lambda <k>/<k^{2}>$), $\lambda=0.8$ in SIS simulations, $\beta$ is the value we used in SIR simulations and $k_{S_{max}}$ is the highest $k$-shell index of the network. We consider only the
  largest connected cluster of the network if the original network is
  disconnected.}
\label{si_table}
\end{table}

f) {\it The Internet at the router level (RL)}. The nodes of the RL
Internet network are the Internet routers.  Two routers are connected if
there exists a physical connection between them. Data have been gathered
from the DIMES project \cite{carmi}. The largest connected component of
the analyzed dataset contains 493312 routers with an average degree of
3.3.

g) {\it The Internet at the autonomous system level (AS)}. The nodes are
autonomous systems which are connected if there exists a physical
connection between them.  An autonomous system is a collection of
connected IP routing prefixes under the control of one or more network
operators that presents a common, clearly defined routing policy to the
Internet. Data have been gathered by the DIMES project \cite{carmi}. The
largest connected component of the AS Internet consists of 20556
autonomous systems with average degree 6.1.

h) {\it Product space of economic goods}. This is the network of proximity
between products according to Ref.~\cite{hidalgo}. We use a proximity threshold 0.3,
and we recover similar results for different thresholds, as well.

We outline some of the basic properties for these networks in
Table~\ref{si_table}.

\section{The $k$-shell decomposition method}
\label{kshell}

\begin{figure}[!ht]
\includegraphics[width=14.0 cm]{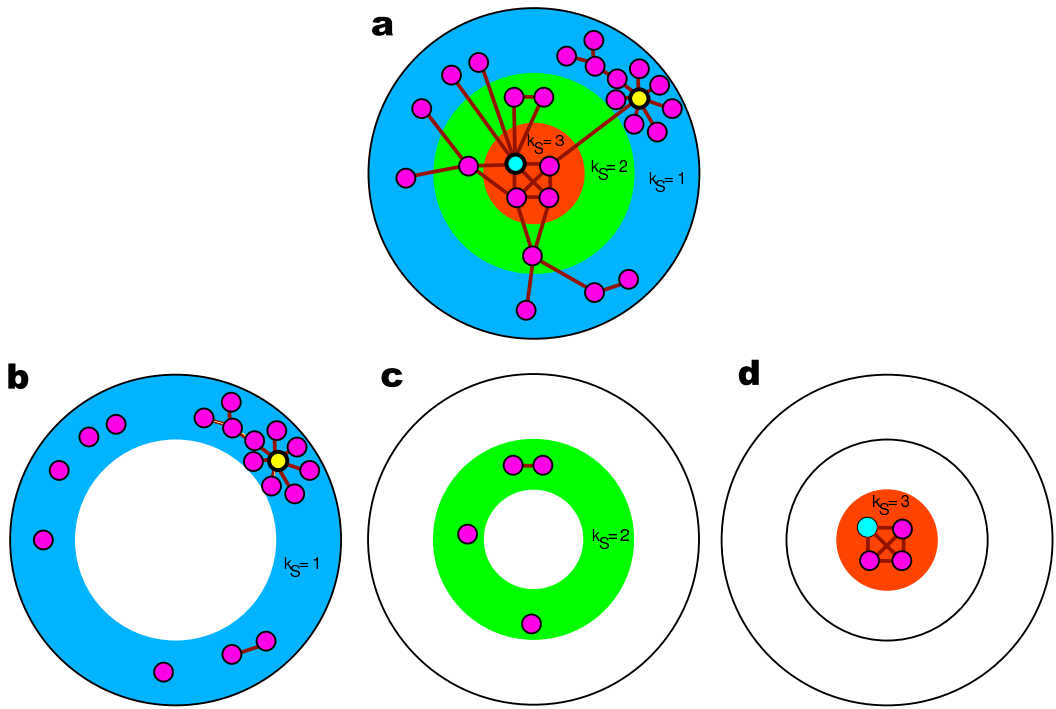}
\caption{{\bf The illustration of the k-shell extraction method.} {\bf
    a,} A schematic network is represented as a set of 3 successively enclosed
  $k$-shells labeled accordingly. {\bf b,} Nodes with edges forming
  $k_S=1$ shell of the network. {\bf c,} Nodes with edges forming
  $k_S=2$ shell of the network. {\bf d,} Nodes with edges forming
  $k_S=3$ shell of the network.} \label{SIdrawing}
\end{figure}

In order to classify the nodes into $k$-shells we employ the $k$-shell
decomposition algorithm. First, we remove all nodes with degree $k$=1.
After this first stage of pruning there may appear new nodes with
$k$=1. We keep on pruning these nodes, as well, until all nodes with
degree $k$=1 are removed. The removed nodes along with the links
connecting them form the $k_{S}=1$ $k$-shell.  Next, we repeat the
pruning process in a similar way for the nodes of degree $k$=2 to
extract the $k_{S}=2$ $k$-shell and subsequently for higher values of
$k$ until all nodes are removed. As a result, the network can be
viewed as a set of adjacent $k$-shells (see Fig.~\ref{SIdrawing}).

The $k$-shell decomposition method assigns a unique $k_S$ value to each node, that
corresponds to the index of the $k$-shell this node belongs to.  The
$k_S$ index provides a different type of information on a node than that
provided by the degree $k$. By definition, a given layer with
index $k_{S}$ can be occupied with nodes of degree $k \geq k_S$. In the
case of random model networks, such as the configurational
model,
there is a strong correlation between $k$ and the
$k_S$ index of a node and, therefore, both quantities provide the same
type of information. Thus, the low-degree nodes are generally in the
periphery, and the high-degree nodes are generally in the innermost
$k$-shells. In real networks, however, this relation is often not true.
In real networks hubs may have very different $k_S$ values and can be
located both in the periphery (yellow node in Fig.~\ref{SIdrawing}) or
in the core (blue node in Fig.~\ref{SIdrawing}) of the network.

\begin{figure}[!ht]
\includegraphics[width=14.0 cm]{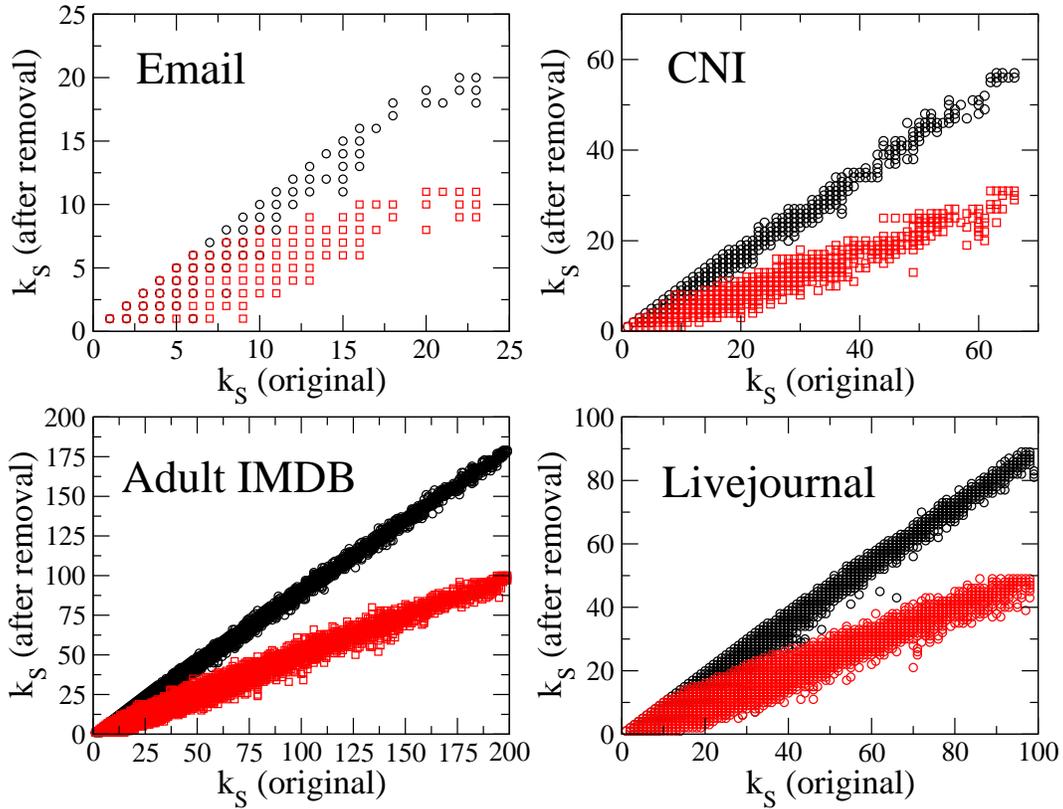}
\caption{{\bf Robustness of $k_S$ under incomplete network information}.
  We randomly remove 10\% of the network links and 50\% of the network links
  (results shown in black and red symbols, respectively). The relative
  ranking of the nodes remains invariant under both removals, for all the
  networks studied: Email, Hospital, Adult IMDB, and Livejournal.com.}
  \label{ks_robustness}
\end{figure}

The assignment of a $k_S$ index to a node is also quite robust.
We have randomly removed 10\% and 50\% of the links in the networks that we
study, simulating thus incomplete information. When we measure the new $k_S$
value for the same nodes in the resulting networks (Fig.~\ref{ks_robustness})
we find that their relative ranking remains the same. We recover a
practically linear dependence on the $k_S$ values of the original and
the incomplete networks, showing that this measure would work equally
well for predicting the spreading efficiency of nodes in a network with
missing information.

\section{Probability and time of infection}
\label{eandt}

We have demonstrated that the location of a node, as described through
the $k_S$ index, is important for the extent of spreading $M_i$ when this node
is the spreading origin. Here, we show that nodes with high $k_S$ are
more probable to be infected during an epidemic outbreak and are infected
earlier than nodes with low $k_S$, when spreading starts at a random node.
We introduce the quantity $E_{i}$, as the probability that a node $i$ is going
to be infected during an epidemic outbreak originating at a random
location, and $T_{i}$, as the average time before node $i$ is infected
during the same process.

\begin{figure}[!ht]
\center
\includegraphics[width=8cm,height=6cm,angle=0]{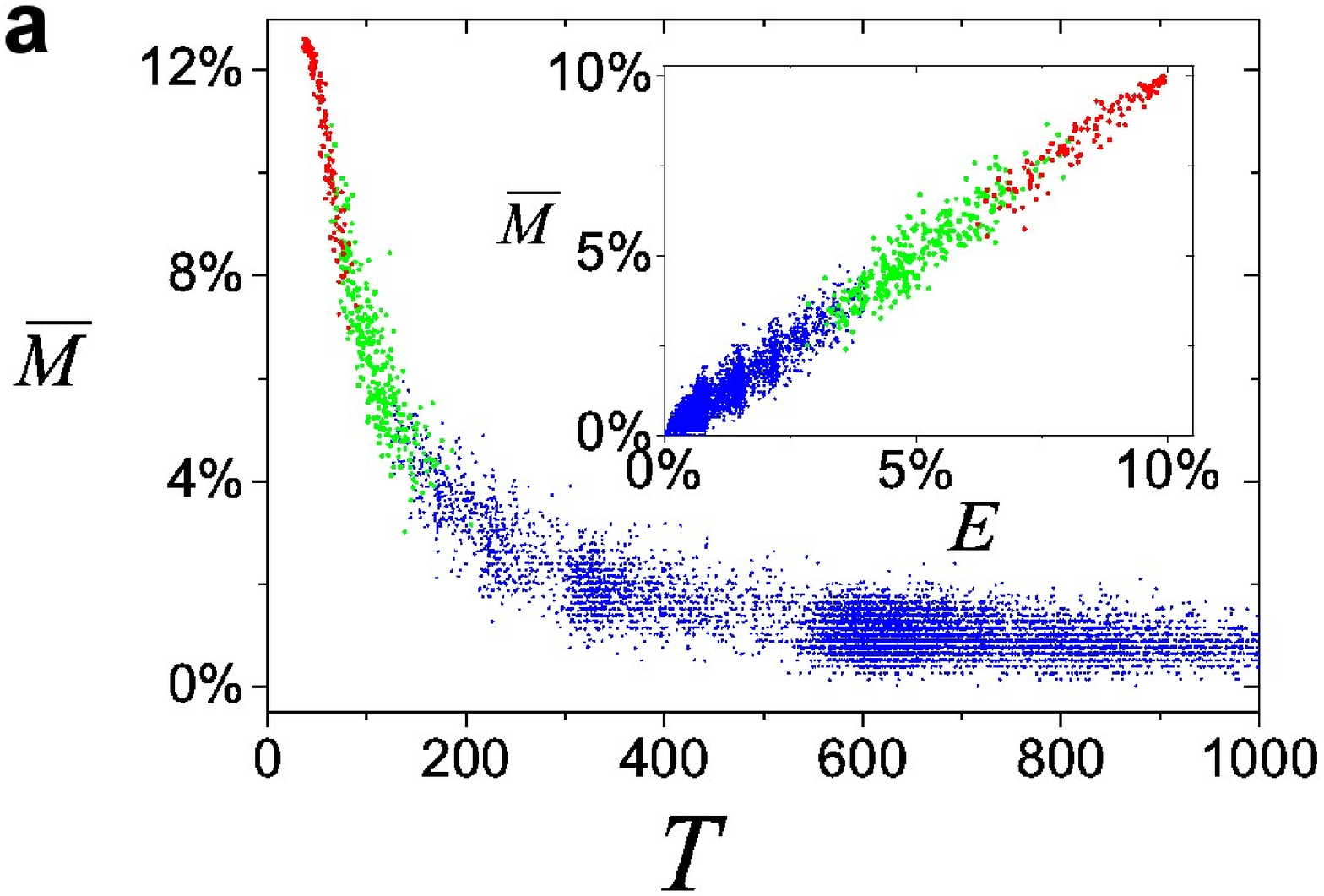}
\includegraphics[width=8cm,height=6cm,angle=0]{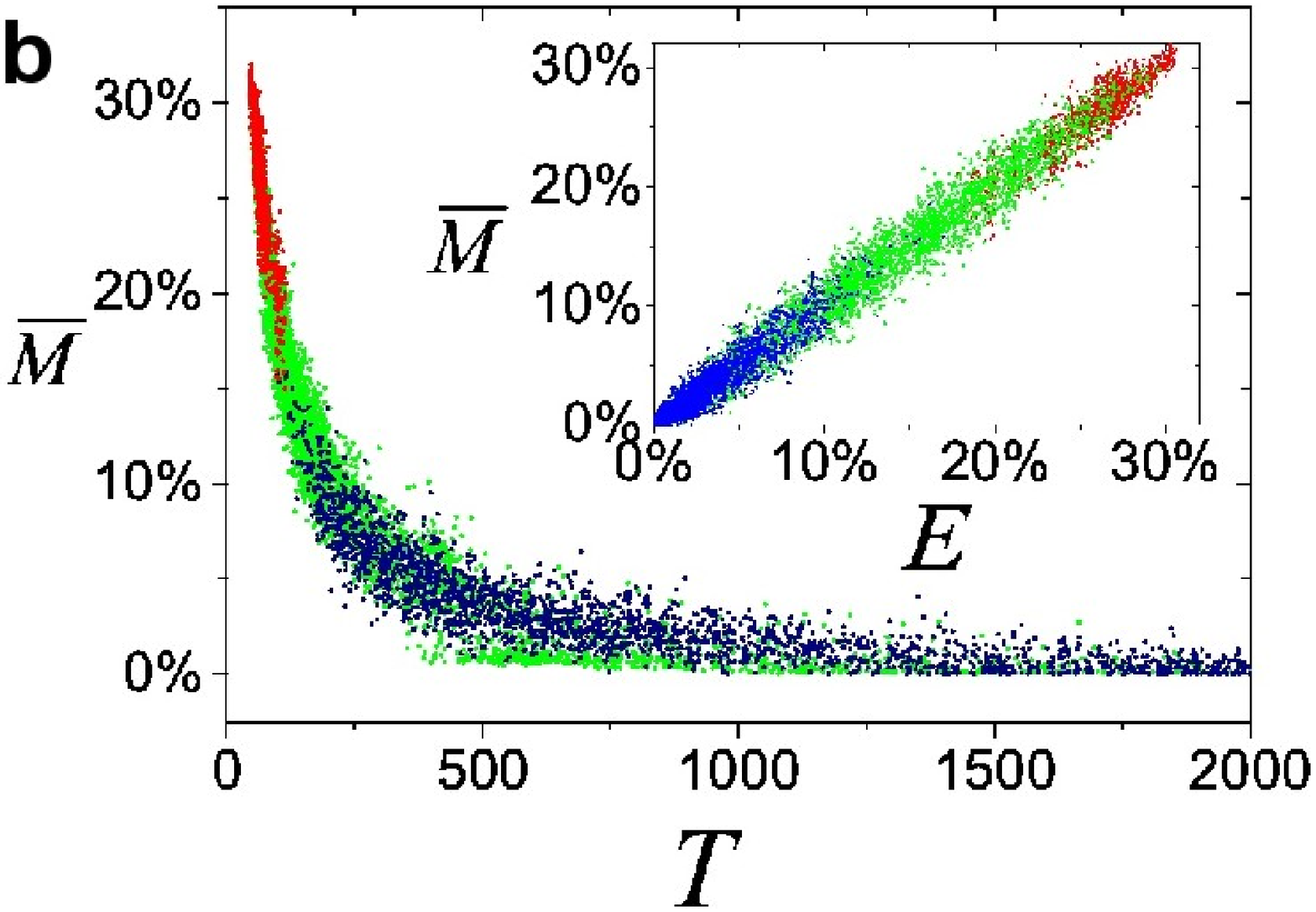}
\includegraphics[width=8cm,height=6cm,angle=0]{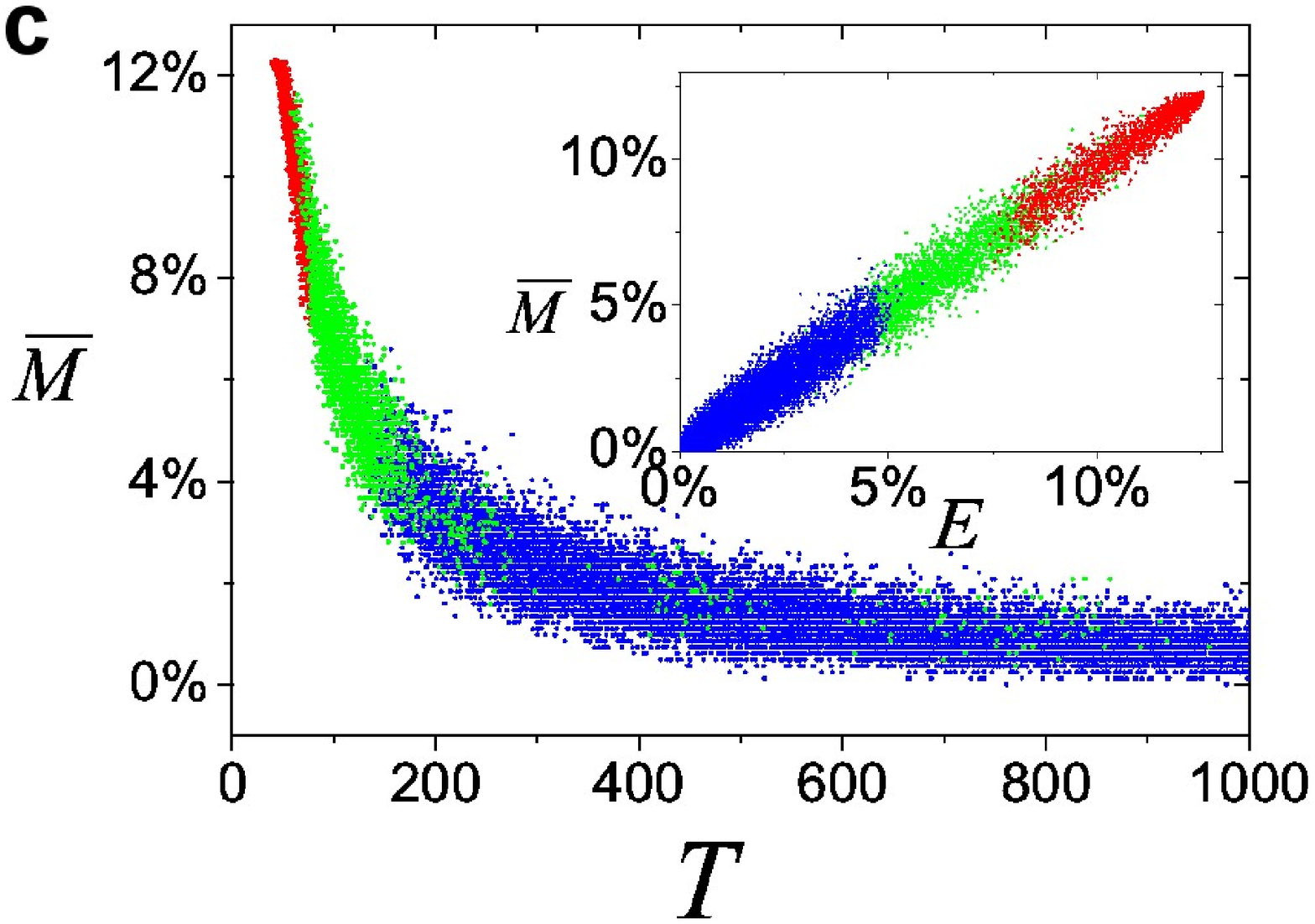}
\includegraphics[width=8cm,height=6cm,angle=0]{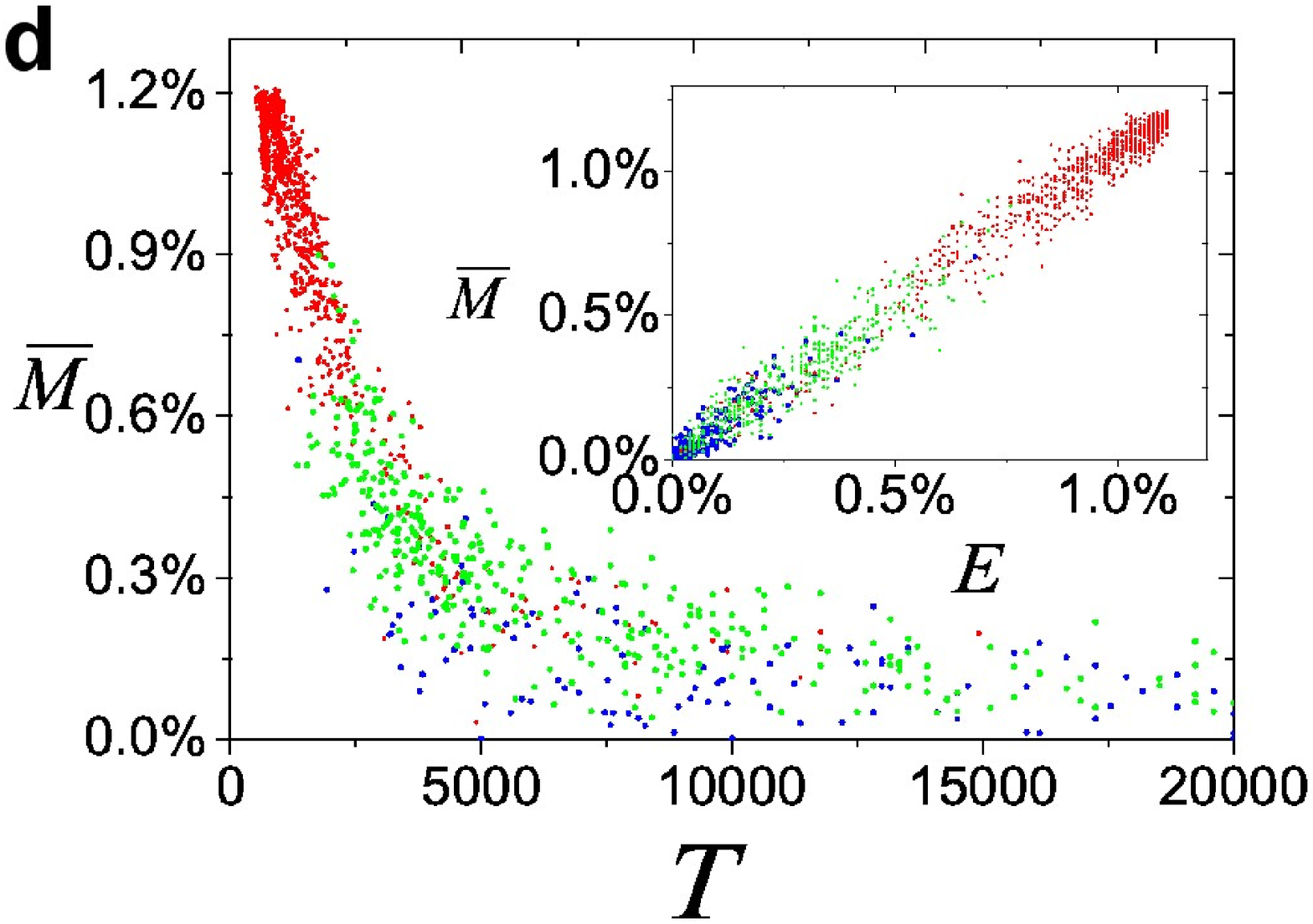}
\vfil\eject
\caption{Cross-plots of $M_i$ as a function of $T_i$, and $M_i$ as a function of $E_i$ (inset) for a) email,
  b) hospital inpatients, c) actor network and d) RL Internet. Every point denotes the corresponding quantities for
  a given node, and the color denotes the $k$-shell index of this node. The $k_S$ values are aggregated and
  highlighted with red (large $k_{S}$ regime), green (intermediate $k_{S}$ regime) and blue (low $k_{S}$ values)
  colors, respectively.  A high level of correlation between $M_i$ and $E_i$
  indicates that the most efficient spreaders (as measured by
  $M_i$) are the most likely to be infected during an epidemic
  outbreak originating at random inpatient in the network. On the other
  hand, the anti-correlation between $M_i$ and $T_i$ indicates
  that the most efficient spreaders are typically infected earlier than
  other nodes during an epidemic outbreak.}
 \label{SItime}
\end{figure}

As shown in Figs.~\ref{SItime}a-d all three quantities that characterize
the role of a node in an epidemics process, $M_i$, $E_i$ and
$T_i$ are strongly correlated. The nodes that are
infected by a given node $i$ form a cluster of size $\overline {M_i}$,
and they are statistically the nodes that can reach $i$ when they act as origins themselves.
Thus, the probability $E_i$ to reach this node in general
is directly proportional to the size $M_i$, as shown in the plots.
The average time $T_i$ to reach a node is inversely proportional to
its spreading efficiency $M_i$, which emphasizes the fact
that these nodes are easily reachable from different network locations.
In conclusion, the nodes with the largest $k_S$ values consistently
a) are infecting larger parts of the network, b) are infected more
frequently, and c) are infected earlier, than nodes with smaller $k_S$
values.

\begin{figure}[!ht]
\center
\includegraphics[width=8cm,height=6cm,angle=0]{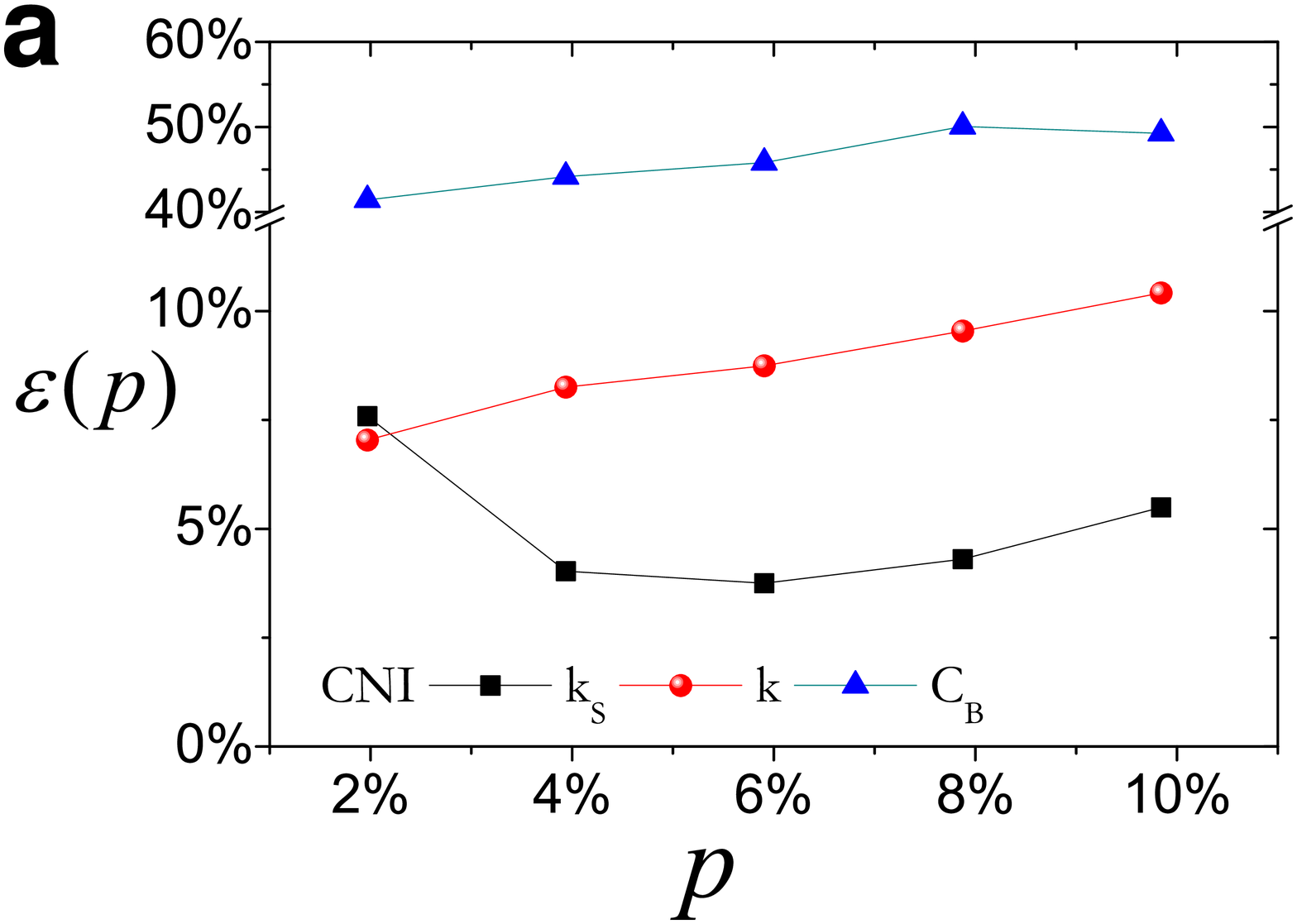}
\includegraphics[width=8cm,height=6cm,angle=0]{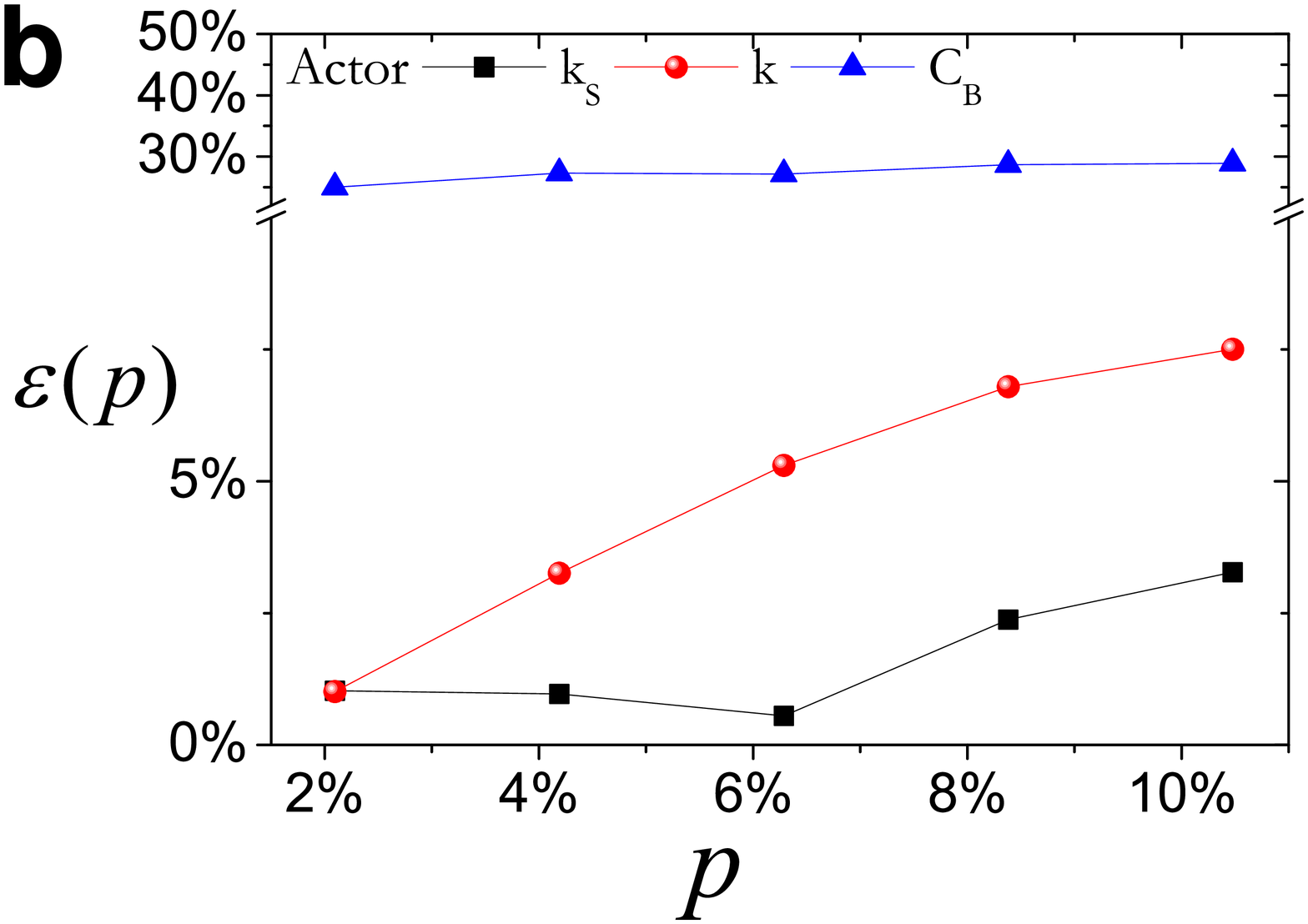}
\includegraphics[width=8cm,height=6cm,angle=0]{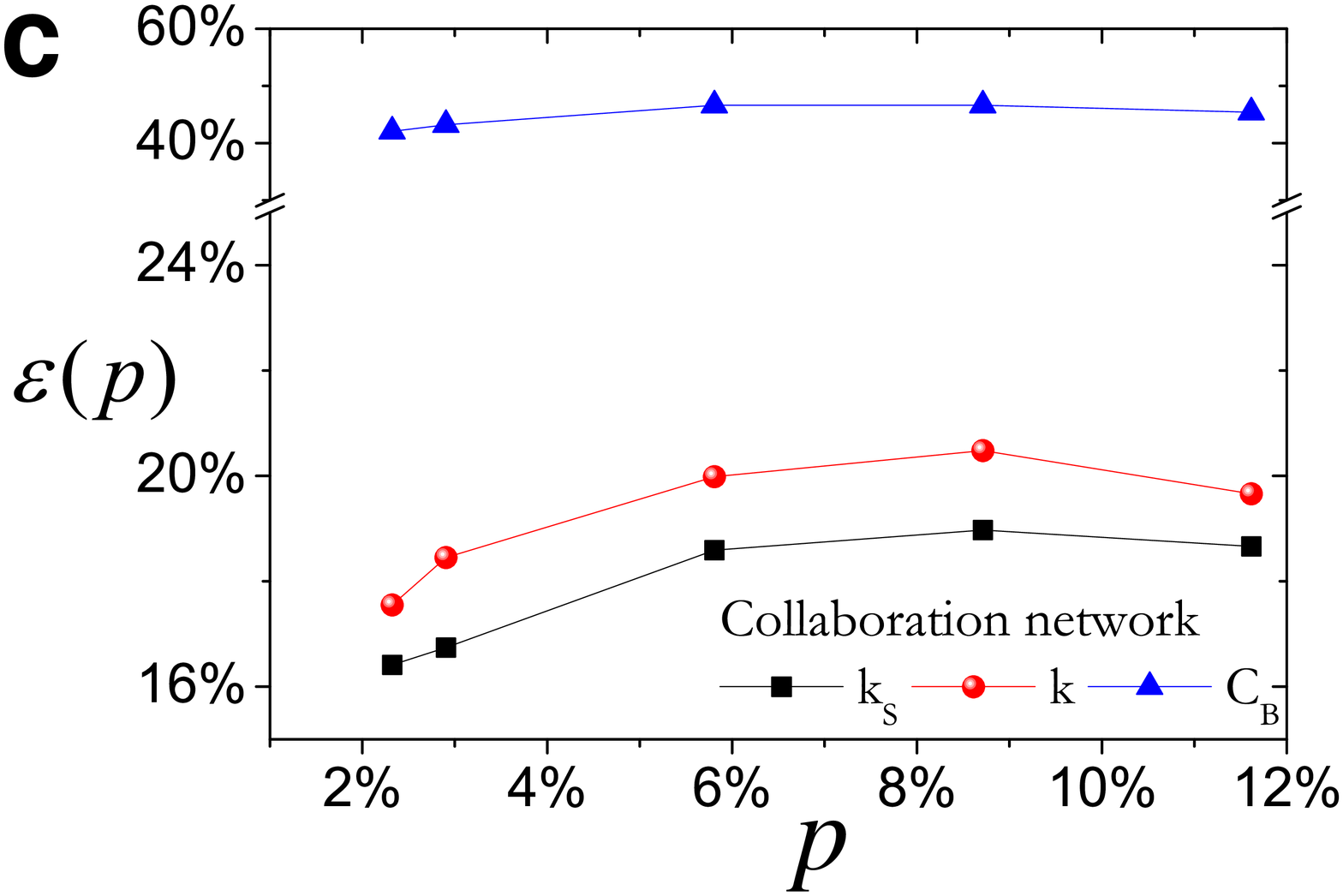}
\includegraphics[width=8cm,height=6cm,angle=0]{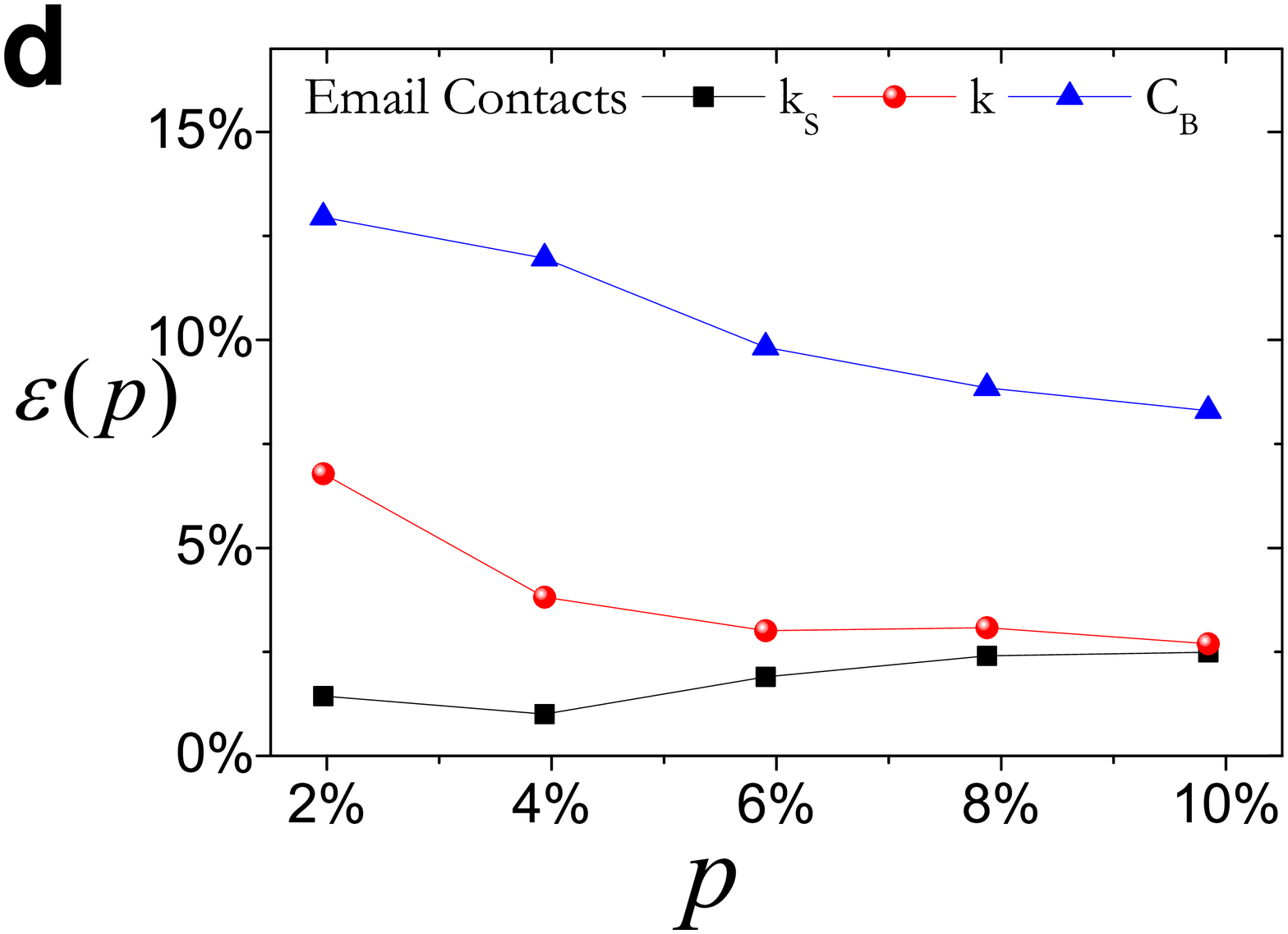}
\vfil\eject
\caption{{\bf The imprecision
  functions $\epsilon(p)$ test the merit of using $k$-shell, $k$ and
  $C_{B}$ to identify the most efficient spreaders in the CNI, actor, collaboration, and email contact
  networks.} The $k$-shell based identification method yields consistently lower imprecision compared to the $k$ and $C_{B}$ based methods.
 \label{SIimprecision}
 }
\end{figure}

\section{The Imprecision Functions}
\label{imprecision}

We quantify the spreading efficiency of an individual origin $i$
through the infected number of nodes $M_i$. In order to compare the
different methods, we rank all network nodes according to their
spreading efficiency, independently of their other properties, and we
consider a fraction $p$ of the most efficient spreaders ($p
\in[0,1]$).  We designate this set by $\Upsilon_{eff}(p)$. Similarly,
we define $\Upsilon_{k_S}(p)$ as the set of individuals with highest
$k$-shell values. In order to assess the merit of using $k$-shell
decomposition to identify the most efficient SIR spreaders one needs
to compare the two sets $\Upsilon_{eff}(p)$ and $\Upsilon_{k_S}(p)$.
In order to consider individual $M_{i}$ values,
we calculate the average $M_{eff}(p)$ and
$M_{k_S}(p)$ values for the sets $\Upsilon_{eff}(p)$ and
$\Upsilon_{k_S}(p)$ respectively: $M_{k_S}(p) \equiv
\sum_{i \in \Upsilon_{k_S}(p)} M_{i} / Np $ and
$M_{eff}(p) \equiv \sum_{i \in \Upsilon_{eff}(p)}
M_{i} / Np $, where $Np$ is the number of nodes that we consider in the comparison.
By definition, $M_{eff}(p) \geq M_{k_S}(p)$, and the equality is only reached if
$\Upsilon_{eff}(p) = \Upsilon_{k_S}(p)$. We assess the imprecision of
$k$-shell identification by calculating the ratio between
$M_{eff}(p)$ and $M_{k_S}(p)$:
\begin{equation}
\epsilon_{k_{S}}(p) \equiv 1 -
\frac{M_{k_S}(p)}{M_{eff}(p)}.
\end{equation}
Similarly, we can define
$\epsilon_{k}(p)$ and $\epsilon_{C_{B}}(p)$:
\begin{equation}
\epsilon_{k}(p) \equiv 1
- \frac{M_{k}(p)}{M_{eff}(p)} \,,\, \epsilon_{C_{B}}(p)
\equiv 1 - \frac{M_{C_{B}}(p)}{M_{eff}(p)}.
\end{equation}
A value for $\epsilon$ close to 0 denotes a very efficient process,
since the nodes that are chosen are practically those that contribute
most to epidemics.  In all cases, the $k_S$ method yields a spreading
that is closer to the optimum than either the degree or the
betweenness centrality. Additionally, this behavior is independent on
the fraction of spreaders $p$ that we consider in each case.

\begin{figure}[!ht]
\includegraphics[width=5.5 cm,height=4.2cm,angle=0]{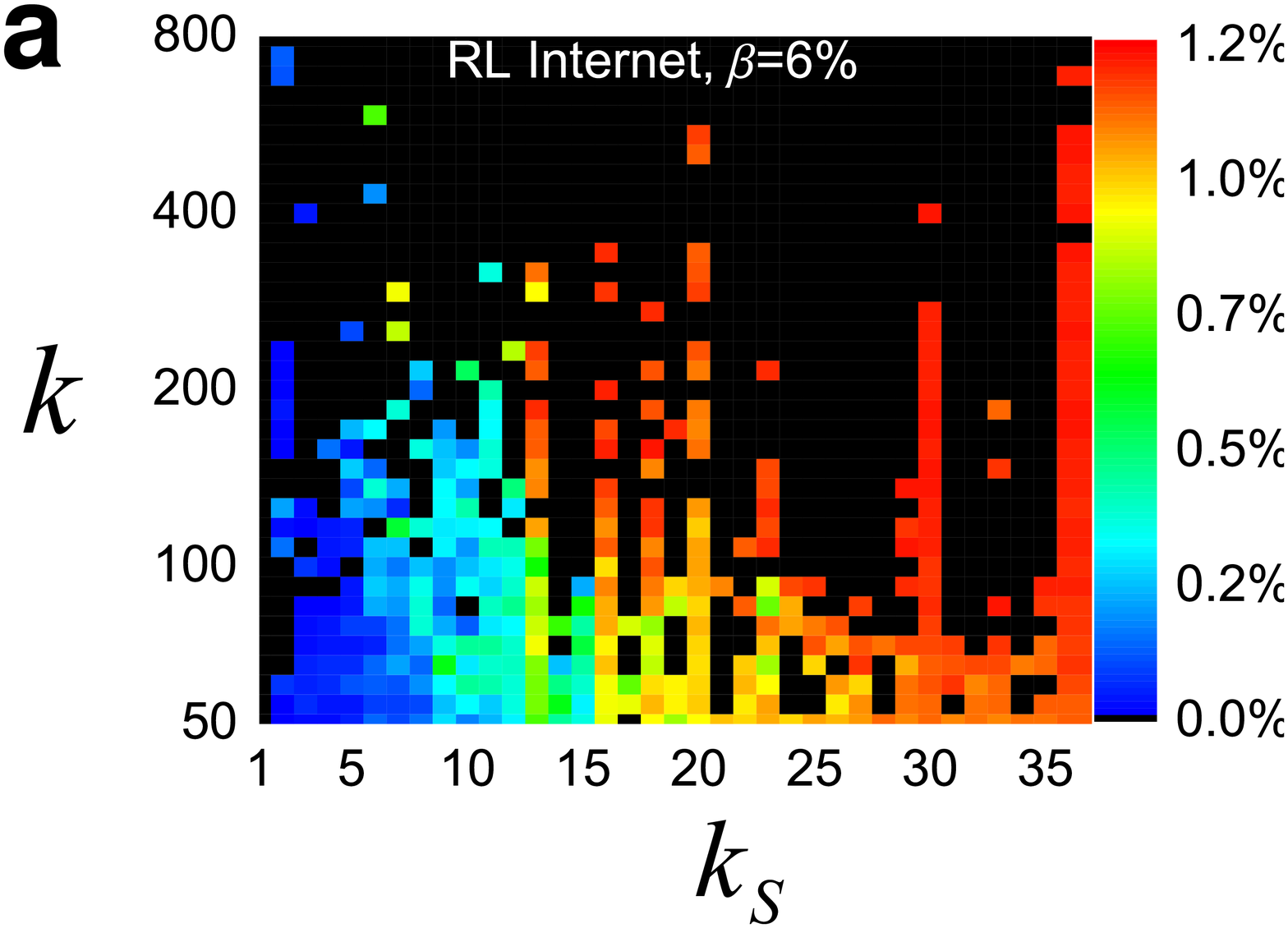}
\includegraphics[width=5.5 cm,height=4.2cm,angle=0]{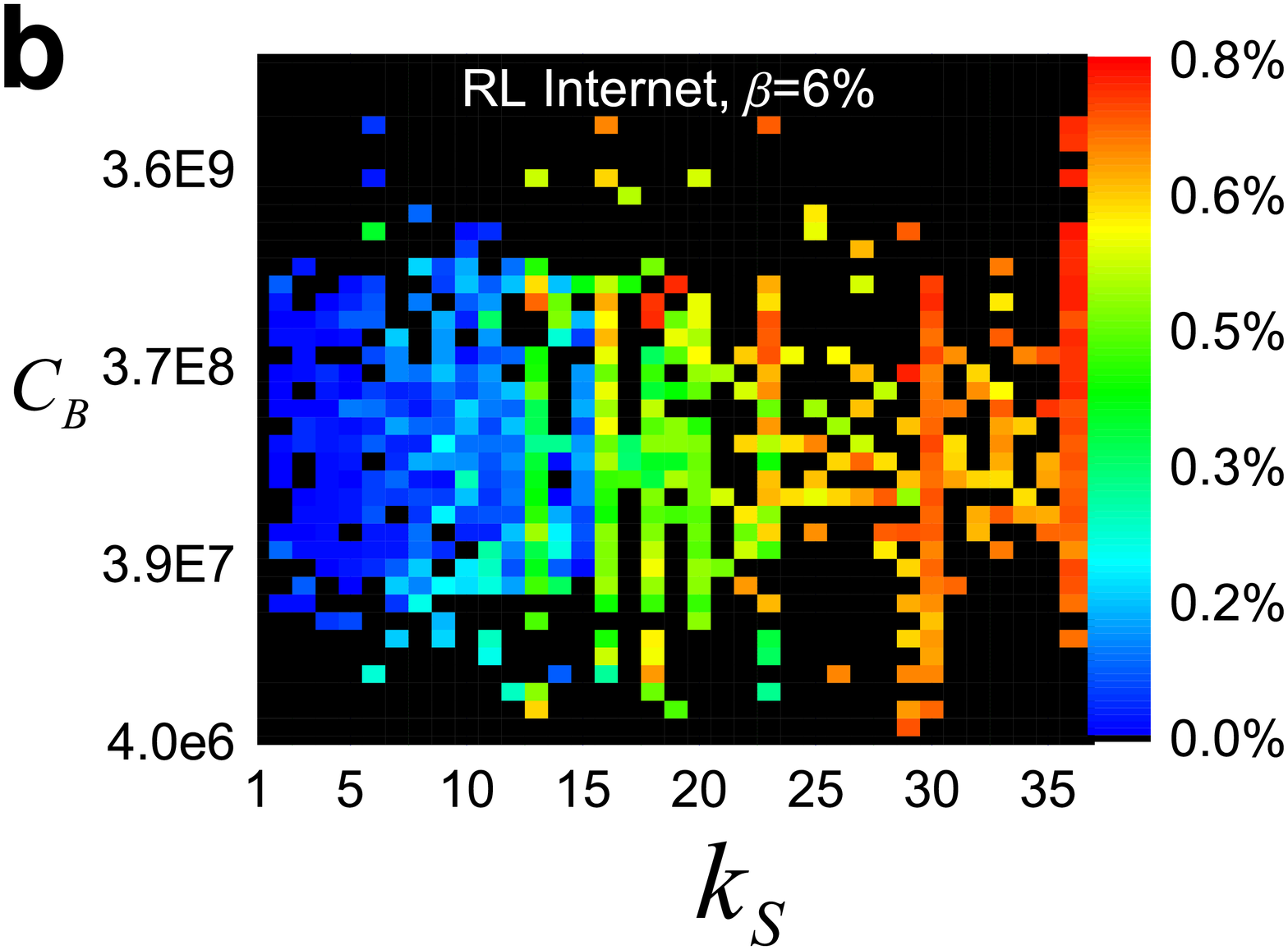}
\includegraphics[width=5.5 cm,height=4.2cm,angle=0]{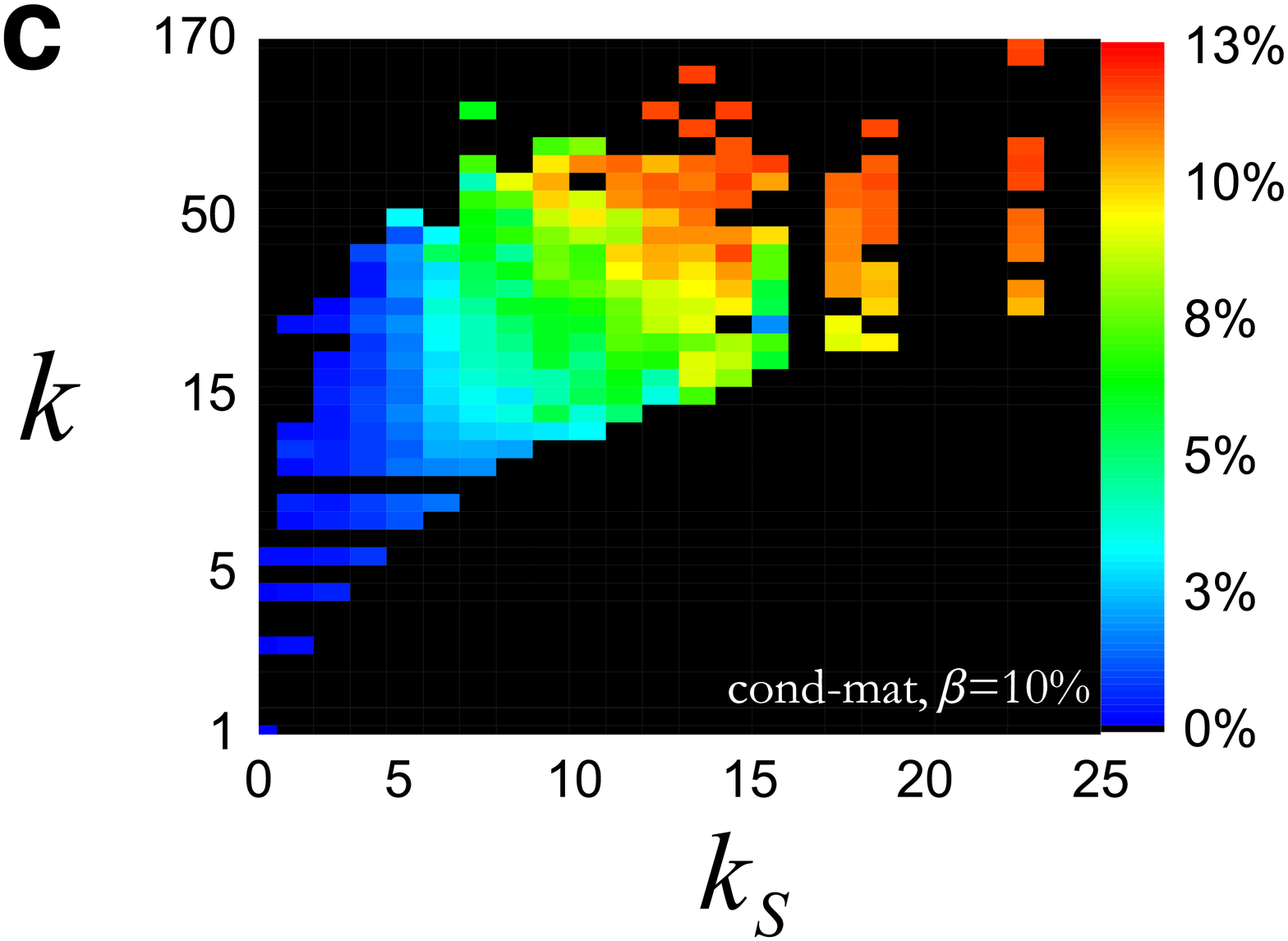}
\includegraphics[width=5.5 cm,height=4.2cm,angle=0]{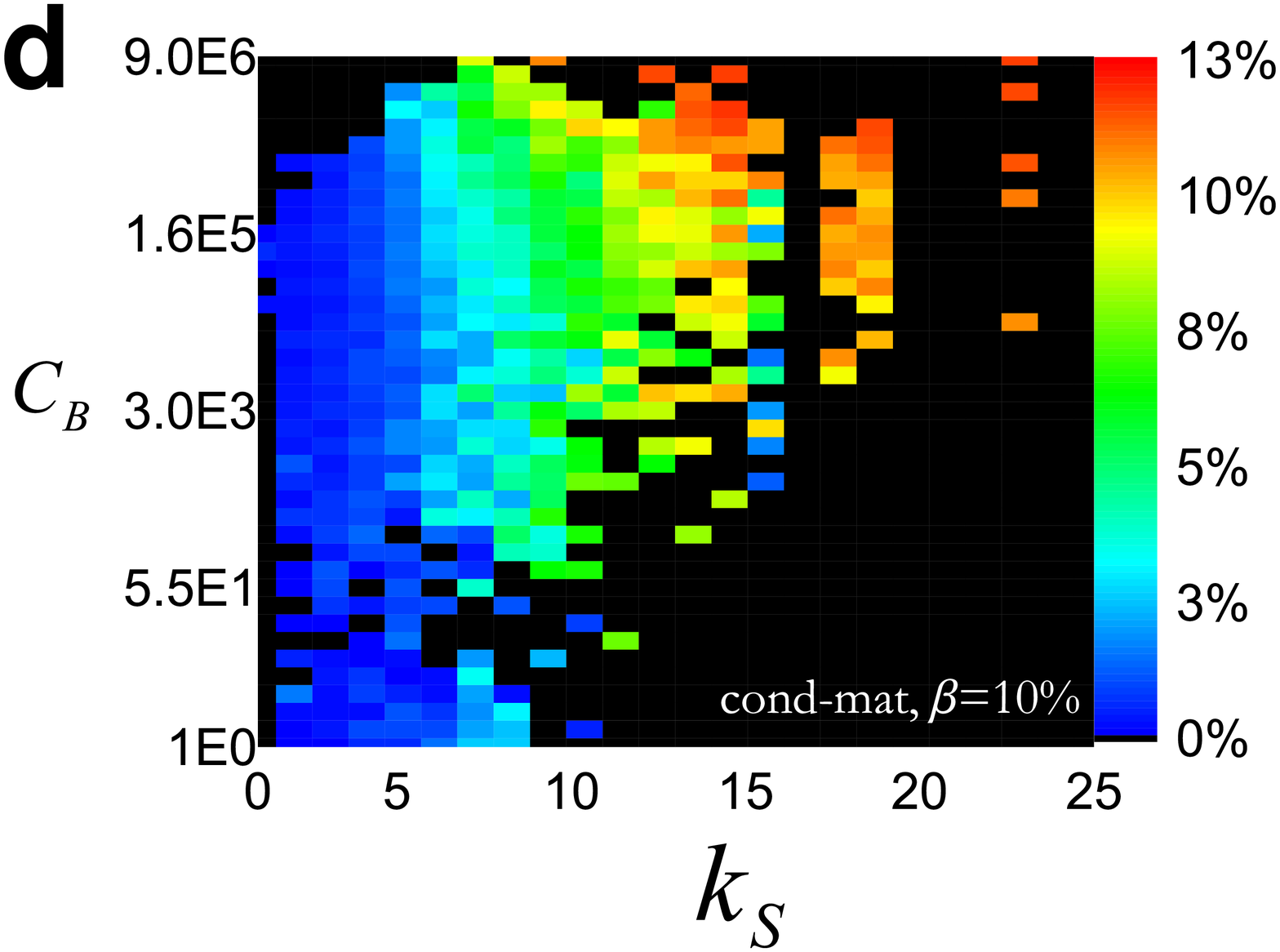}
\caption{ {\bf The shell index $k_S$ predicts the outcome of spreading
    more reliably than the degree $k$ or the betweenness centrality
    $C_B$.} The networks that were analyzed are: ({\bf a, b}) the RL
  Internet and ({\bf c, d}) the collaboration network.  {\bf a} and
  {\bf c}, The average infected size $M(k_S,k)$ as a function of
  ($k_S$,$k$) values of the infection origin nodes.  {\bf b} and {\bf
    d}, The average infected size $M(k_S,C_B)$ as a function of
  ($k_S$,$C_{B}$) values of the infection origin
  nodes.} \label{SIincidence}
\end{figure}

\section{SIR Spreading Efficiency}
\label{other}

In the main text we present results for $M(k_S,k)$ for the email
network, the CNI, the actor network and the Livejournal network.
Here, we present additional results of the $k$-shell analysis of the
Internet at the Router Level (RL) and the scientific collaboration network.
Figure~\ref{SIincidence} shows the results for $M(k_S,k)$ and
$M(k_S,C_B)$. The conclusion on the spreading importance of high $k_S$
nodes is exactly the same as for the social networks in the main text.

\begin{figure}[!th]
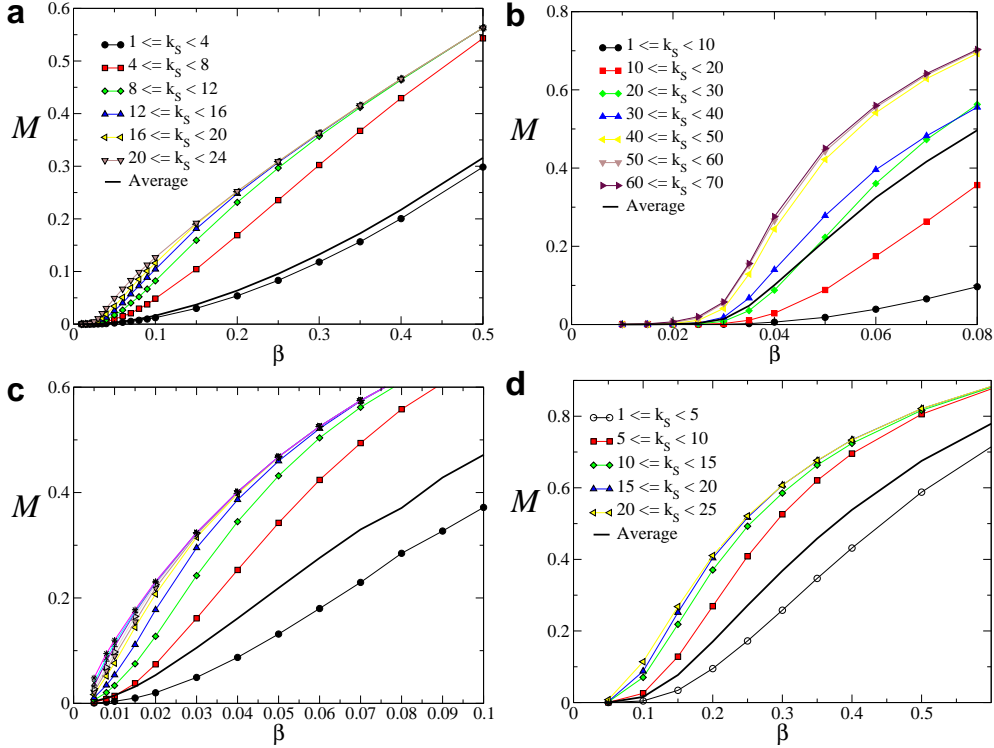

\includegraphics[width=6.5 cm]{SI_fig10a.eps}
\includegraphics[width=6.5 cm]{SI_fig10b.eps}
\includegraphics[width=6.5 cm]{SI_fig10c.eps}
\includegraphics[width=6.5 cm]{SI_fig10d.eps}
\caption{ {\bf The infected percentage is always higher in higher $k$-shells,
independently of the infection probability $\beta$.}
Nodes are grouped according to their $k$-shell and we calculate the average
infected percentage for each group as a function of $\beta$. The solid lines
correspond to the grand average over all nodes acting as spreading origins.
The networks that were analyzed are: {\bf a,} the email network,
   {\bf b,} the CNI, {\bf c,} the adult IMDB actors network, and {\bf d,}
   the cond-mat collaboration network.
 } \label{SIbeta_variation}
\end{figure}

The results on the nodes efficiency are not significantly influenced by the
choice of the infected probability value, $\beta$. In
Fig.~\ref{SIbeta_variation} we present the infected percentage $M$ for different
networks, as an average over nodes that belong in the same $k_S$ range,
for different $\beta$ values.
The nodes in higher $k$-shells are consistently reaching a larger fraction
of the network. Our main interest is in the $\beta$ range where we are above
the critical point, $\langle M \rangle>0$, but the average infection reaches
a finite but small fraction, in the range of 1-20\%. When the average spreading
is even larger, nodes of lower $k$-shells can become efficient too, because in
this case there is a high probability to reach the `core' of the network,
and this would enable the spreading to extend over an even larger part of the network.

\begin{figure}[!ht]
\includegraphics[width=7.0 cm]{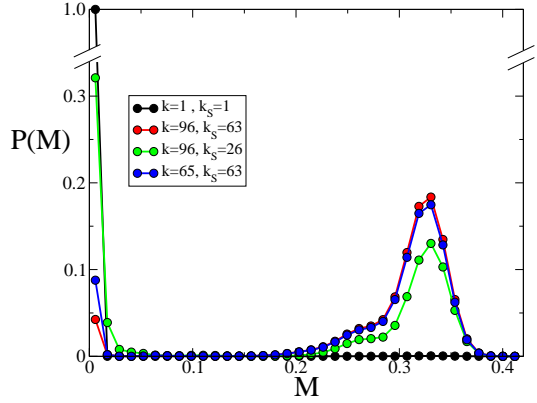}
\caption{{\bf Distribution of spreading based on individual origins.}
  The probability distribution $P(M)$ of the infected percentage for the contact
  network of inpatients, when the epidemic starts at four nodes of different properties.
  The infection probability is $\beta=4\%$, which is above the critical threshold.
  All distributions exhibit two peaks at similar ranges every time, i.e. around
  $M=0$ (epidemics dies very fast) and $M\simeq33$\%. However, the intensity of each
  peak differs, and in higher $k$-shells the majority of the realizations result in
  large infections, compared to the much higher ratio of zero-spreading realizations
  for origins of small $k_S$ values.
  } \label{SIcomp_distr}
\end{figure}

For $\beta$ values in this `intermediate' range, the distribution $P(M)$ of the
infected percentage $M$ is composed by two well-defined peaks (Fig.~\ref{SIcomp_distr}).
The first is at $M=0$ and corresponds to those instances where the infection dies
within the first few infection steps. The second peak is at a finite fraction $M$,
and it seems to be at the same point for all origins. However, the intensity of each
peak strongly differs, depending on the $k_S$ value of the origin. For the higher $k_S$
value in the plot, the stronger peak is at the non-zero value, and very few realizations
end up at $M=0$ even for smaller degrees. On the contrary, an origin with larger degree $k$,
but smaller $k_S$ value results in a stronger peak at $M=0$. These distributions converge
quite well, and we can expect that nodes with small $k_S$ will in general result in a higher
peak at $M=0$. The above means that if an infection can reach a critical mass of nodes
then it will eventually cover a significant part of the network. The low $k$-shell nodes
cannot reach this critical mass so that the infection dies at the early stages, resulting
to the strong peak at $M=0$. On the contrary, the neighborhood of high $k$-shell nodes
is favorable for sustaining an infection at early stages, allowing
the system to reach this critical mass.

\clearpage

\begin{figure}[!ht]
\includegraphics[width=16cm]{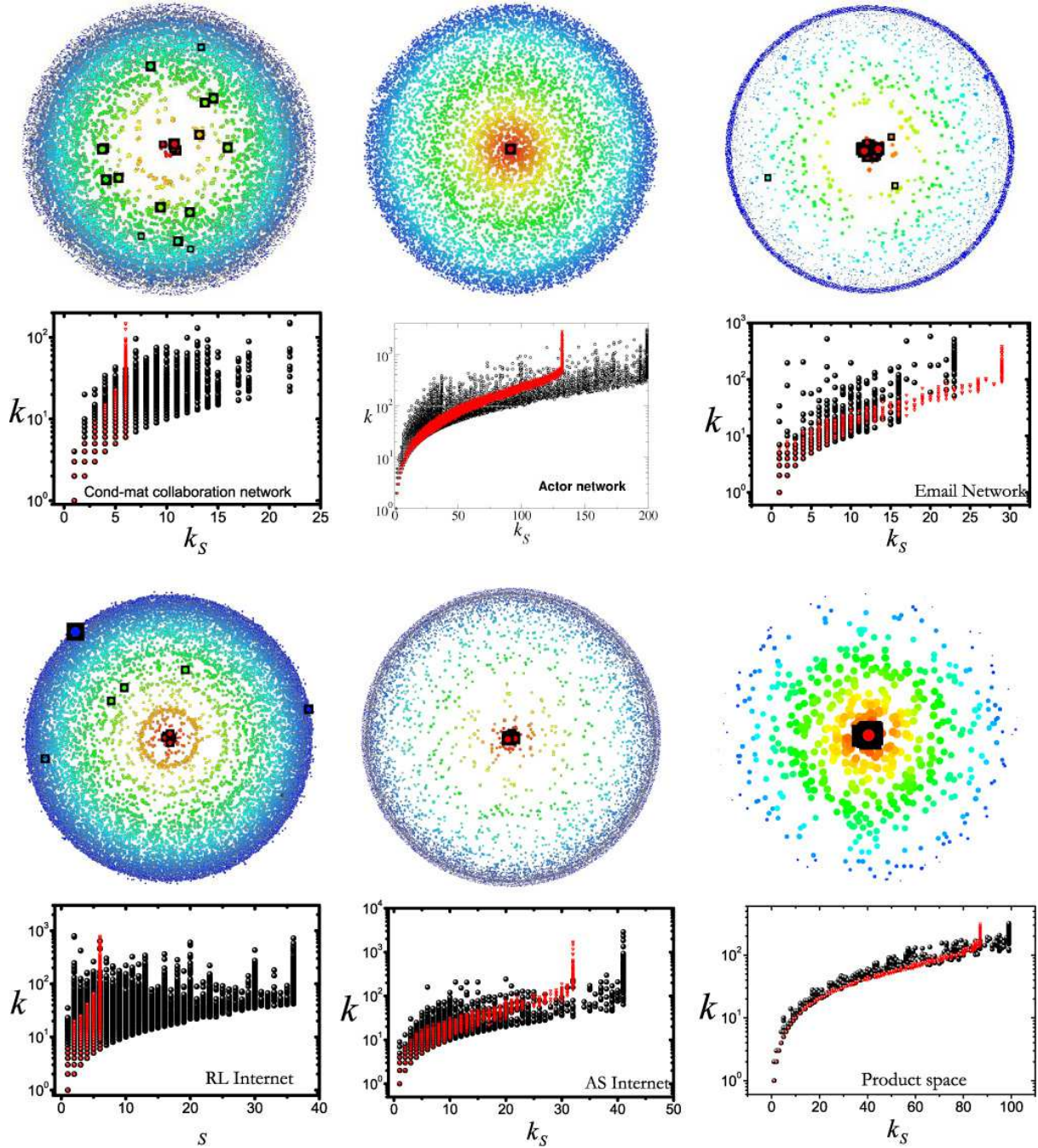}
\caption{{\bf $k$-shell structure of the analyzed networks.}  ({\bf
    Top row}): Visualization of the $k$-shell structure.  We represent
  networks as sets of concentric circles of nodes, each one
  corresponding to the particular $k$-shell, with low $k_S$ values in
  the periphery and large $k_S$ values towards the center of the
  network.  The size of each visualized node is proportional to the
  logarithm of its degree value.  We highlight the 25 highest degree
  nodes with black squares. Many of the hubs are found in outer
  layers.  ({\bf Bottom row}): Scatter plots of node degree $k$ as a
  function of its $k$-shell index $k_S$ for the original networks
  (black symbols) and the degree-preserving randomized version of the
  networks (red symbols).  The networks correspond to: the cond-mat
  collaboration network, the actor network, the email contact network,
  the RL Internet, the AS Internet, and the Product Space network.}
\label{SIscatter}
\end{figure}

\clearpage

We also highlight the location of the 25 largest hubs in the $k$-shell
structure of the studied networks.  Fig.~\ref{SIscatter} shows the
results for the collaboration, actor, email, RL Internet, AS Internet,
and Product space networks.  High-degree nodes in most of the studied
networks are scattered at different $k$-shells: the high-$k$
nodes appear both in the periphery (starting as low as $k_S=1$) and in
the network center (large $k_S$ value).  In certain cases, such as in
the actors network, the largest hubs are located in the highest $k_S$
layers.  The relation of $k_S$ and $k$ in the AS Internet and the
product space is strongly monotonic, and there are very few nodes
where $k_S$ is large or small compared to the degree $k$. This is a
typical behavior for random networks, and the structure of these two
networks is significantly close to their randomized counterparts. In
these cases, choosing a node based on its degree or its $k$-shell
index does not make a difference, since they practically lead to the
same nodes.

\begin{figure}[!ht]
\includegraphics[width=15.0 cm,]{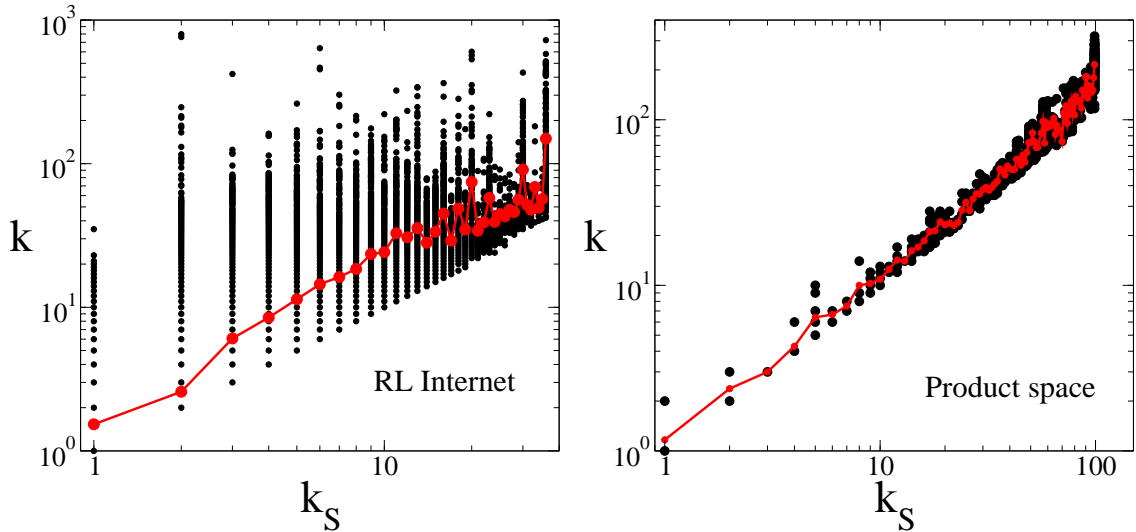}
\caption{{\bf Deviations from the average behavior highlight the importance
of the $k$-shell structure.} The average degree (red symbols) for a given $k_S$
index follows roughly a power-law dependence, as a function of $k_S$. The deviation
from this behavior can be significant, e.g. in RL internet, or negligible, as e.g. in
the product space network.}
\label{SIpowerlaw}
\end{figure}

It is clear that the assortative behavior in a network can influence
the extent to which hubs will appear in the periphery or in the core
of a network. In principle, in a highly disassortative network we expect
more hubs in the periphery, due to their tendency to connect to low-degree nodes.
However, even in assortative networks it is possible that some hubs may
still belong to low $k$-shells, so that the $k_S$ value will appropriately
rank even these exceptions. The average degree of the nodes in a specific
shell follows roughly a power law with $k_S$ (Fig.~\ref{SIpowerlaw}).
The deviations from this average behavior emphasize the importance
of spreaders within the core of the network having high values of $k_S$
and potentially smaller degrees, than those with high $k$ and low
$k_S$ values.

The complex organization of the nodes in the $k$-shells is highlighted
when we randomly rewire the links in the networks, yet preserving the
nodes degree.  This rewiring `restores' all the hubs to the innermost
$k$-shell of the system and imposes a strict hierarchy of nodes in
terms of both $k$ and $k_S$.  The bottom row of plots in
Fig.~\ref{SIscatter} shows the scatter-plots of degree $k$ as a
function of $k$-shell index $k_{S}$ for every node in the network.  In
all cases, a monotonic relation of $k$ vs $k_S$ is followed in the
'rewired' networks (red symbols), where now all the hubs appear in the
highest $k$-shell) as opposed to the weak correlation between $k$ and
$k_S$ in the original networks (shown in black).

\section{Rewiring highlights the importance of $k$-shell}
\label{rewiring}

\begin{figure}[!ht]
\includegraphics[width=10.0 cm,angle=0]{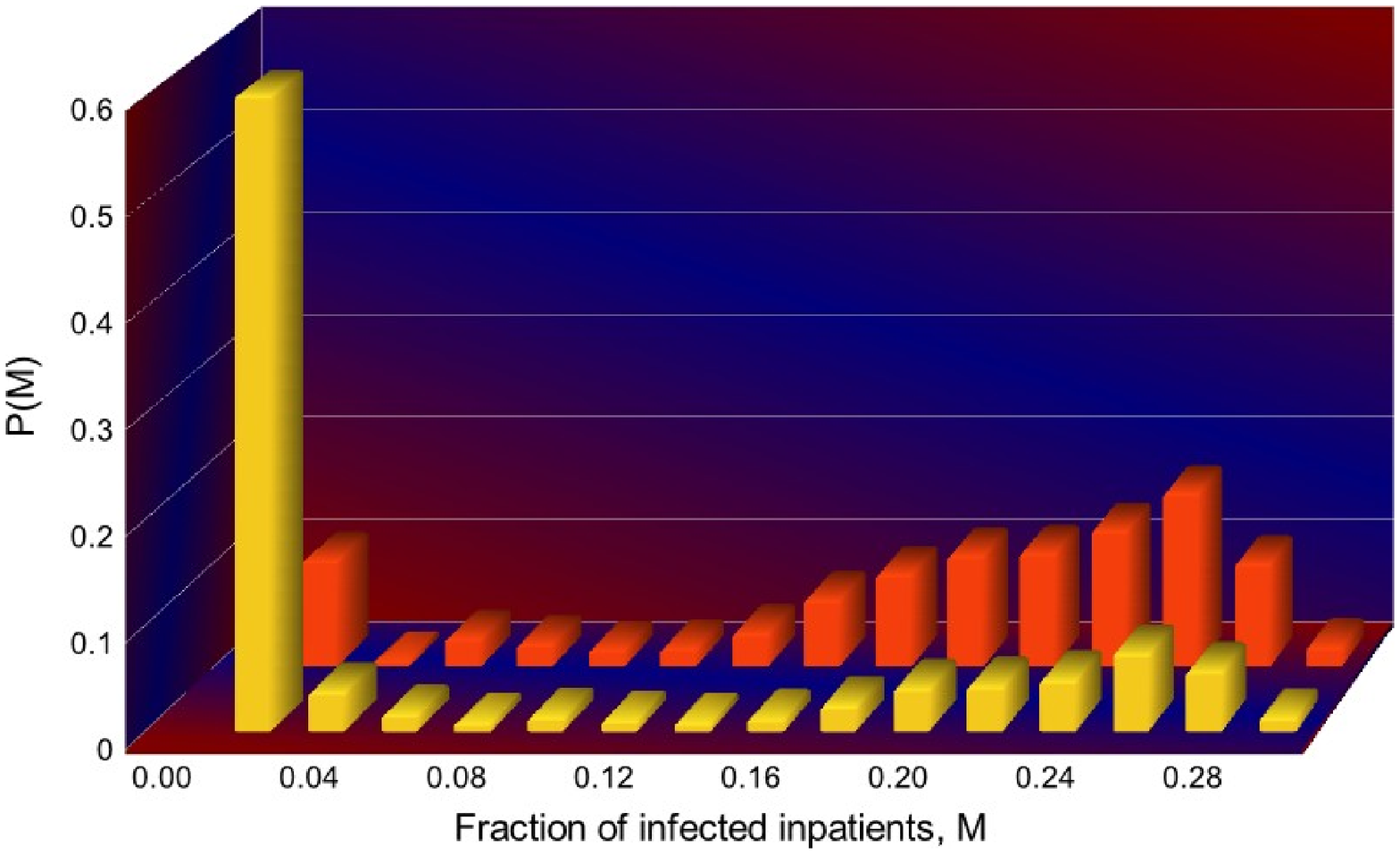}
\includegraphics[width=10.0 cm,angle=0]{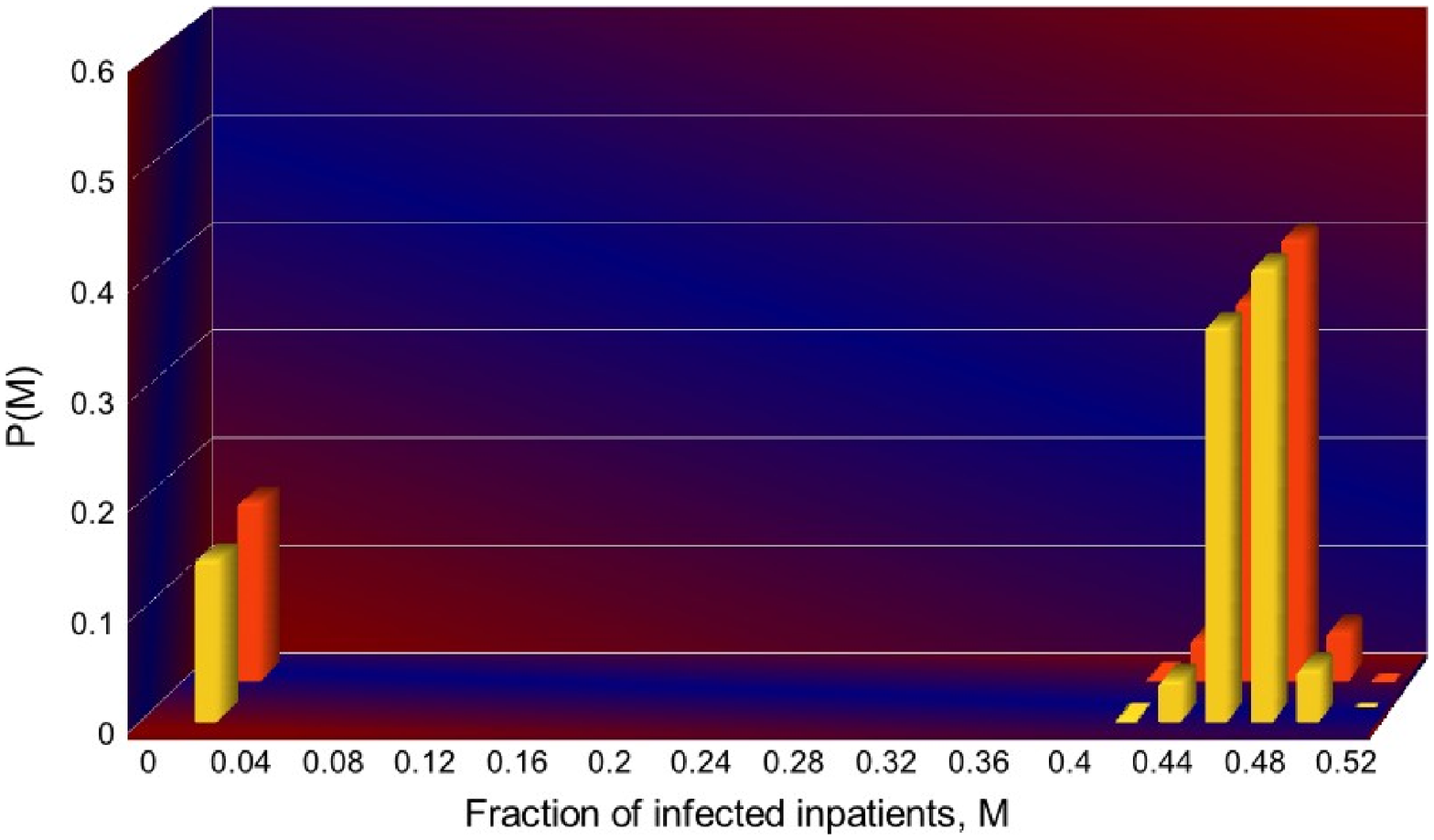}
\caption{{\bf Why the hubs may not be good spreaders.} The probability
  distribution $P(M|i)$ of the infected percentage for the contact
  network of inpatients, when the epidemic starts at two of the origin hubs in Fig.~\ref{histograms}
  $i={A,B}$ with the same degree ($k=96$), but different $k_S$ values
  ($k_S=63$ and $k_S=26$, respectively).
  In each histogram, we use 1000 random realizations of the simulation,
  starting an SIR epidemic from the same given origin $i$. Despite the fact that the two origins
  of the epidemic spreading have the same degree, the two
  histograms present a radically different character.  In one case (red
  histogram), the hub infects up to 30\% of the
  population, while most of the spreading attempts from the other hub
 (yellow histogram) practically cannot propagate the
  infection at all. The importance of the organization of the network is
  highlighted when we randomly rewire the network (preserving the same
  degree for all nodes).  In this case both distributions $P(M|A)$ and
  $P(M|B)$ coincide and both hubs contribute equally to spreading.
  Notice also that spreading in the rewired network extends over a much
  larger size of the population.
  } \label{SIhistograms}
\end{figure}

In Figs.~1a and 1b of the main text we show that the extent of
infection can be remarkably different, although we start from two
origins with similar degree.  The importance of the structure in the
dynamics of spreading can be highlighted if we randomly rewire the
network.  During this process the original degrees of all nodes are
preserved, but random neighbors are chosen for each node, destroying
thus any correlations and any patterns in the local connectivity.  We
denote by $P(M|i)$ the probability that a percentage $M$ of the total
population will be infected if a disease originates on node $i$. In
Figs.~1a,b of the main text and in Fig.\ref{SIhistograms}a we show
that two nodes \#1 and \#2 with similar degree may yield markedly
different distributions $P (M |1)$ and $P (M |2)$.
After rewiring, these distributions become practically
indistinguishable (see Fig.~\ref{SIhistograms}b).

\section{Virus Persistence in SIS}
\label{persistance}

Many infectious diseases, including most sexually transmitted
infections, do not confer immunity after infection, so that they cannot
be described via the SIR model. These cases are better simulated
through the SIS epidemic model \cite{hethcote00}. The dynamics of SIS
epidemics is different, since the number of infected nodes eventually
reaches a dynamic equilibrium ``endemic'' state at which exactly as
many infectious individuals become susceptible as susceptible nodes
become infected \cite{hethcote00}. The quantity characterizing the
role of nodes in SIS spreading is the persistence, $\rho_{i}(t)$,
defined as the probability that node $i$ is infected at time $t$
\cite{satorras01}. In an endemic SIS state, which is reached
asymptotically, $\rho_{i}$ becomes independent of $t$. The persistence
$\rho$ has been shown to be higher in hubs which are reinfected
frequently due to the large number of their neighbors
\cite{satorras01,satorras02,dezso02}. To uncover the role of $k$-shell
layers in SIS spreading we use the joint persistence function
\begin{equation}
\rho(k_S,k)\equiv\sum_{i\in\Upsilon(k_S,k)}\frac{\rho_i}{N(k_S,k)}.
\end{equation}

\begin{figure}[!ht]
\center
\includegraphics[height=22cm,angle=0]{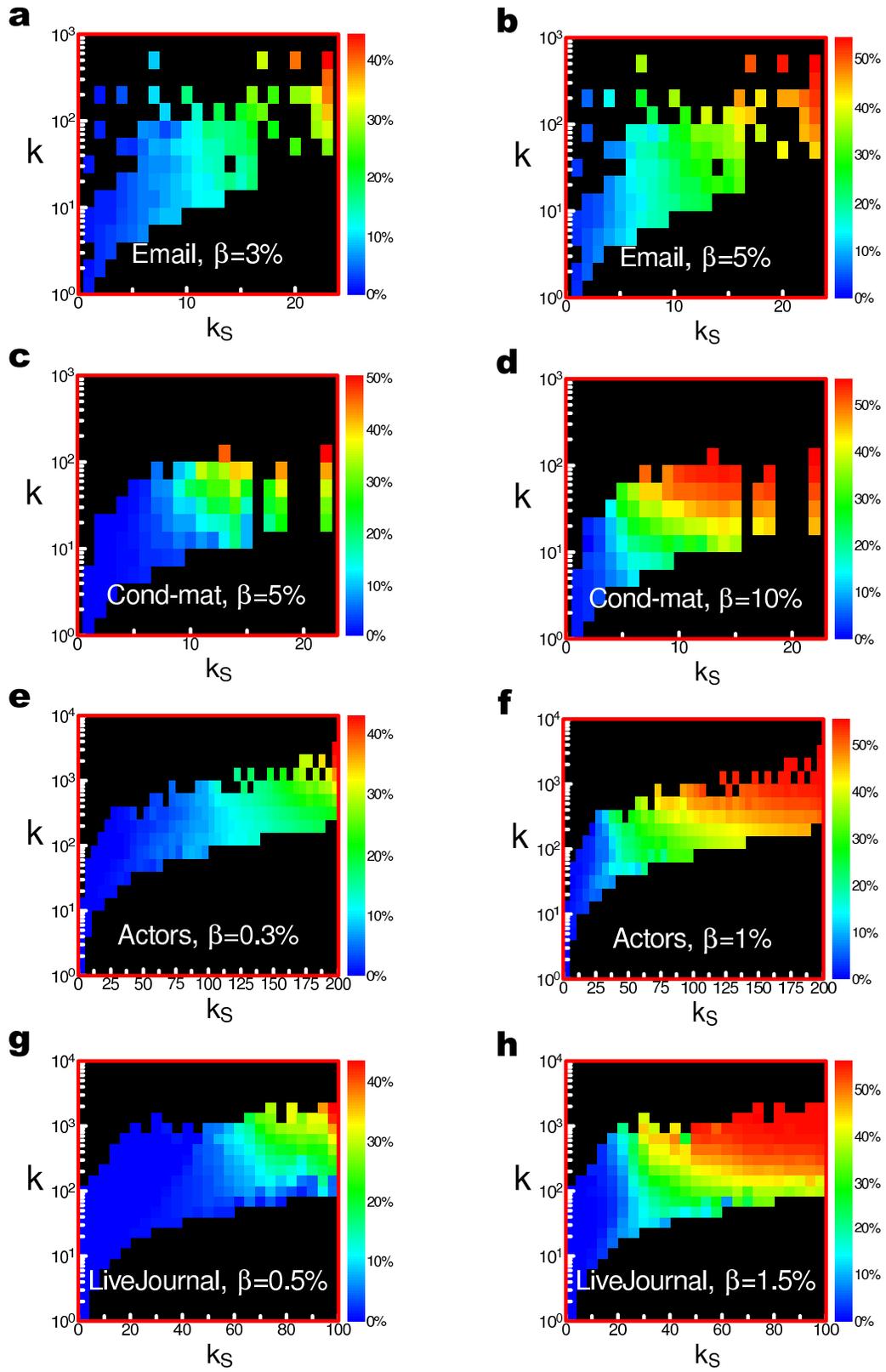}
\vfil\eject
\caption{SIS maps}
\label{SImaps}
\end{figure}

Here we present results for the virus persistence in the Actor,
Collaboration, Email and RL Internet Networks.  Similar to
Fig.~\ref{sis}, we depict $\rho(k_S,k)$ in both supercritical ($\beta
> \beta_{c}$) and subcritical ($\beta < \beta_{c}$)
regimes, where $\beta_c$ is the critical threshold.
In the supercritical regime, $\rho(k,k_S)$ increases with
both $k$ and $k_S$, with maximum values corresponding to hubs in the
innermost layers (see Fig.~\ref{SImaps}). As depicted in
Fig.~\ref{SImaps}, in the subcritical regime, viruses persist only in
the highest $k_S$ layers, while the probability of finding an
infected node in low $k$-shells is negligible.

\begin{figure}[!ht]
\center
\includegraphics[width=8cm,height=6cm,angle=0]{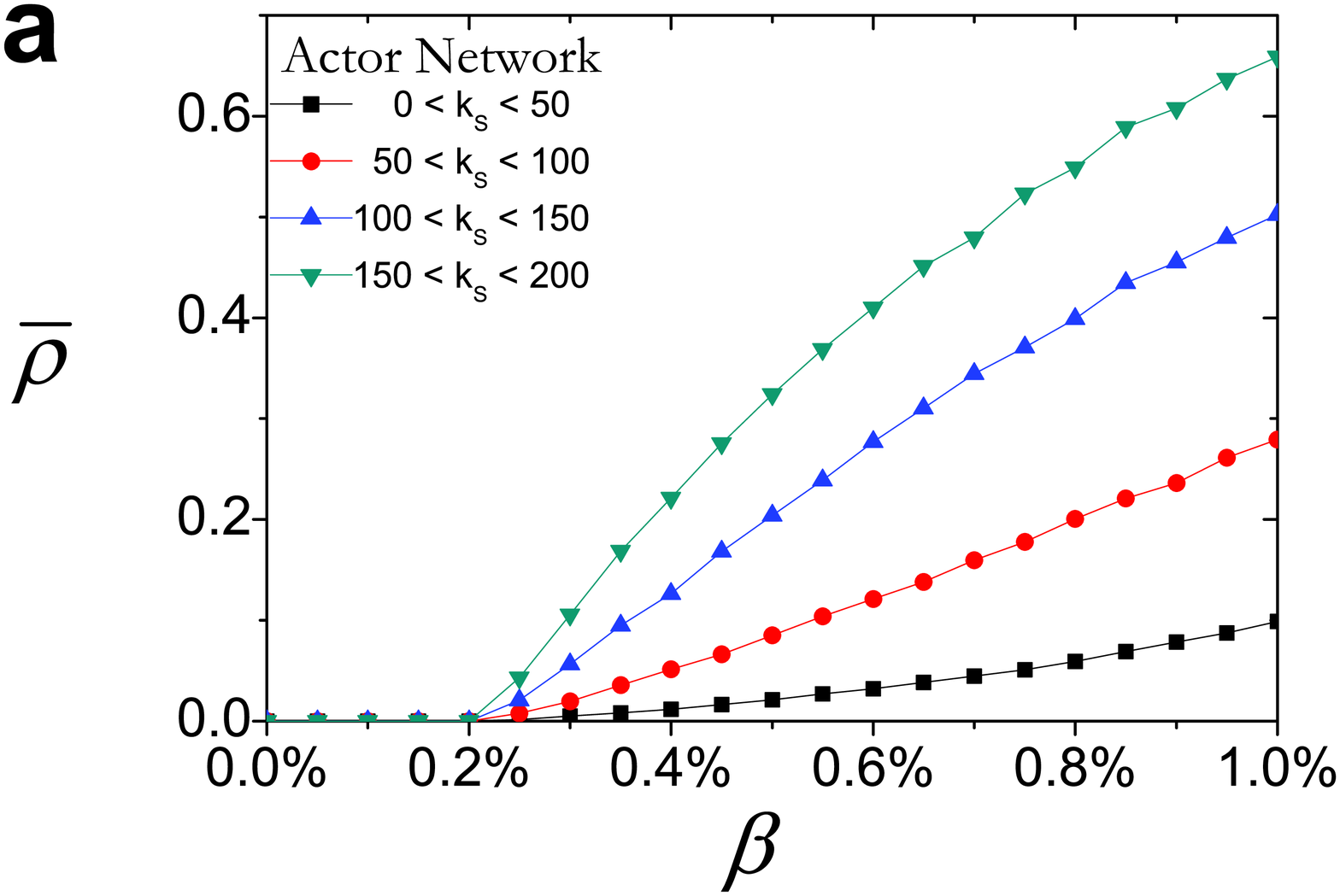}
\includegraphics[width=8cm,height=6cm,angle=0]{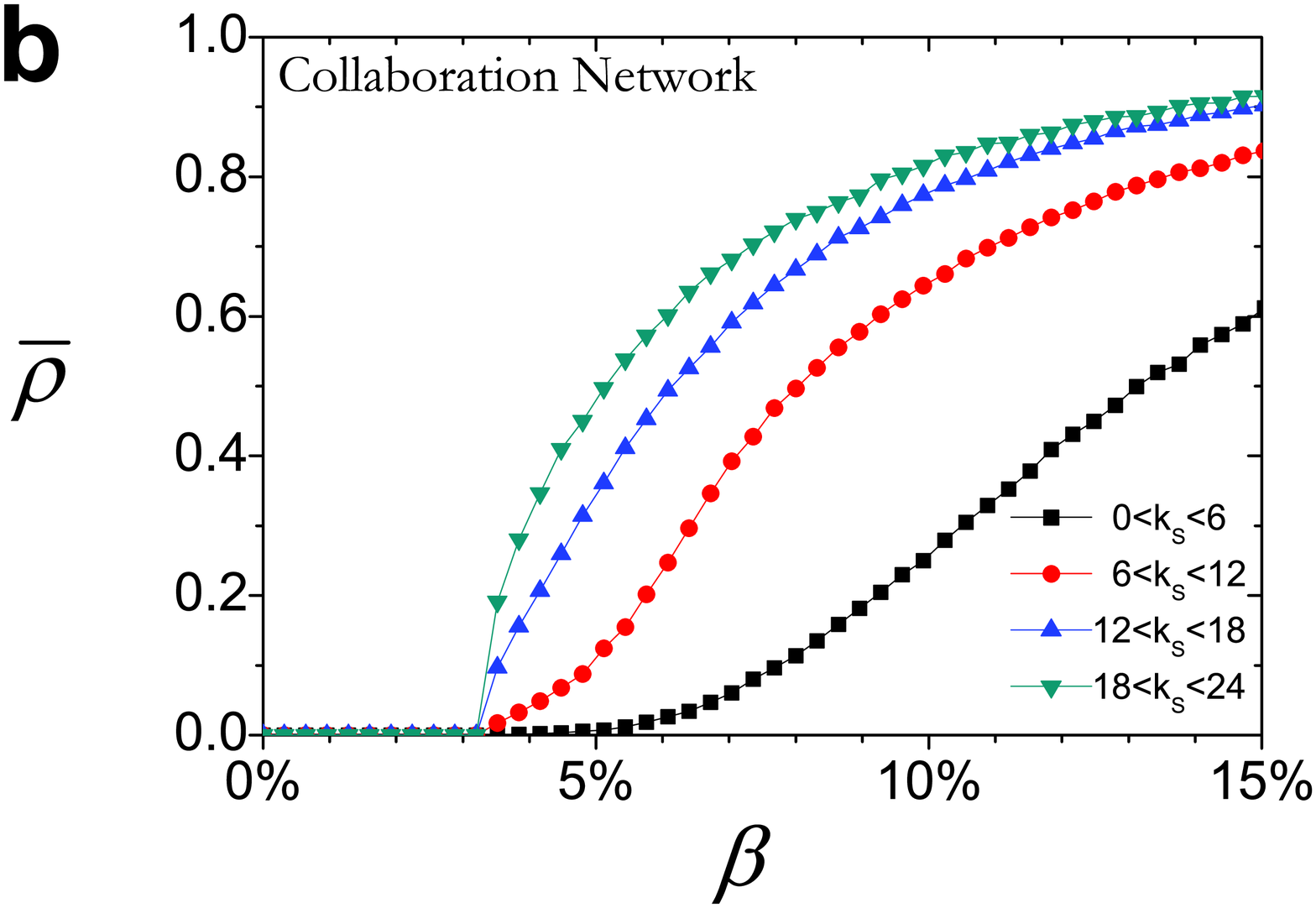}
\includegraphics[width=8cm,height=6cm,angle=0]{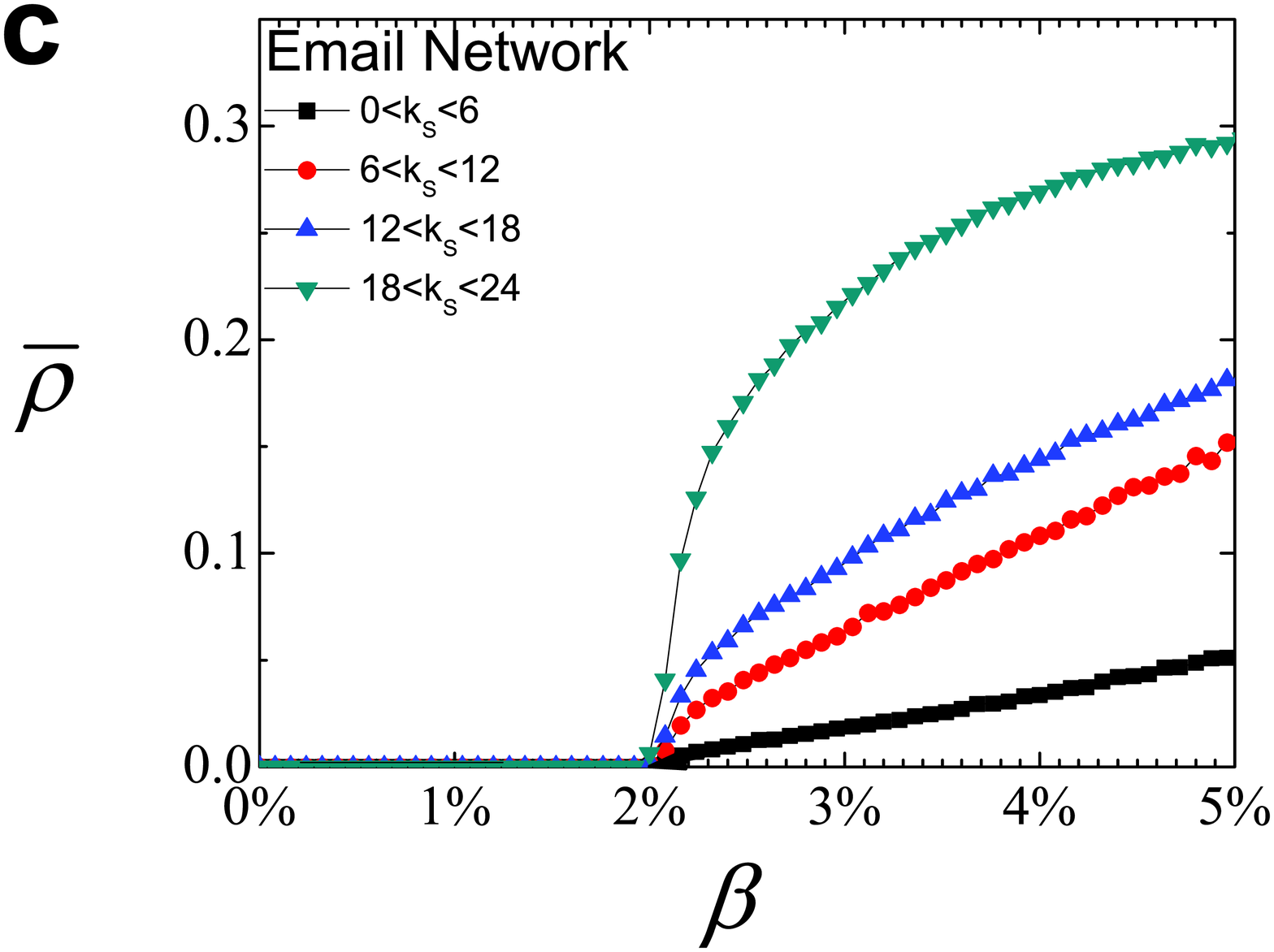}
\includegraphics[width=8cm,height=6cm,angle=0]{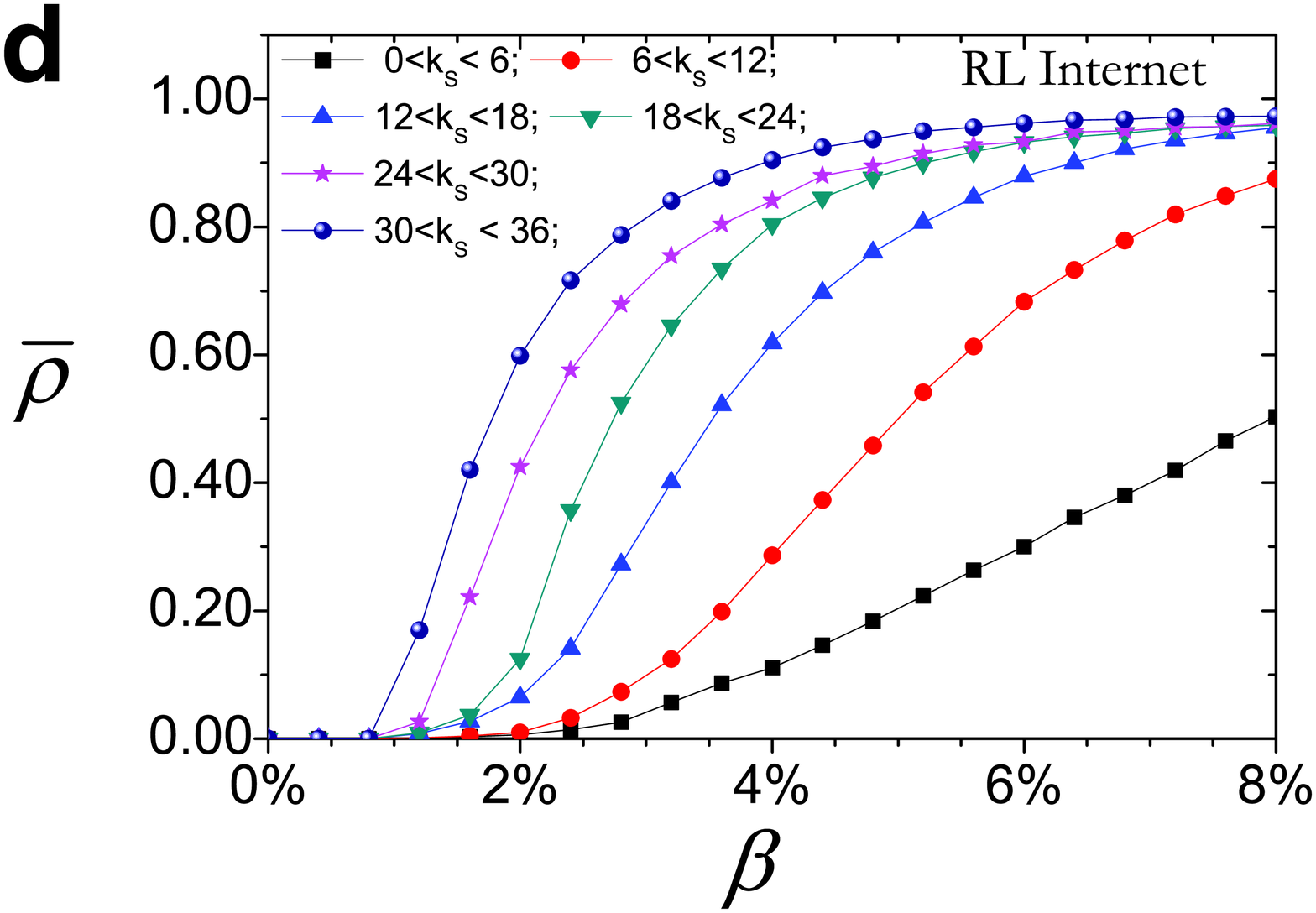}
\vfil\eject
\caption
{
{\bf How average SIS persistence in different $k$-shells depends on virus contagiousness.}
For every network we randomly sample several groups of nodes based on $k$-shell index (as described in SI).
We plot the average virus persistence $ \overline{\rho}$ for every group of nodes as a function of $\beta$ for
the Email, Actor, Collaboration and RL Internet networks. Virus persistence is higher for nodes located in
higher $k$-shells.
}
\label{SIbeta}
\end{figure}

In order to determine in the above networks the actual epidemic
threshold $\beta_c$ we study the behavior of SIS spreading over a wide
range of $\beta$ values. In order to
highlight the role of $k$-shells in spreading, we organize
several groups of nodes based on the $k_S$ layers of each
network. Every such group comprises approximately 100 randomly chosen
nodes with the corresponding $k$-shell indices. In order to achieve
similar average degree in each of the groups, we pick nodes with
uniform probability based on their degree.  As shown in
Fig.~\ref{SIbeta}, virus persistence is consistently higher in the
inner $k$-shells for all values of $\beta$. Moreover, we find
substantially lower epidemic thresholds than in the random cases
$\beta_c<\beta^{\rm rand}_c$ in all
considered networks except for the Email Contact network.

The results of Figs.~\ref{SImaps} and \ref{SIbeta} suggest that the
observed persistence of a virus is due to the dense sub-network formed
by nodes in the innermost $k$-shell, which helps the virus to
consistently survive locally in this area.  Indeed, the innermost
layers can be regarded as a small subgraph exclusively
consisting of hubs. By definition, all nodes in this innermost
$k$-shell will have degrees $k \geq k_{S_{max}}$. Therefore, as a simple
approximation, one can regard the innermost core of a network as a
regular graph consisting of nodes with the same degree
$k=k_{S_{max}}$.

The mean-field solution of the SIS spreading in a regular graph can be
found, for instance in Ref.~\cite{satorras02}. We reproduce this
solution below for the sake of convenience.

The master equation describing the time evolution at a mean-field
level of the average density of infected individuals $\rho(t)$:
\begin{equation}
\frac{d \rho(t)}{d t} = -\rho(t) + \beta k \rho(t) (1-\rho(t)),
\label{SIregular}
\end{equation}
where $k$ is the degree of all nodes in the regular graph. The first
term on the right hand side of Eq.~(\ref{SIregular}) accounts for
infected nodes becoming healthy. The second term on the right hand
side of Eq.~(\ref{SIregular}) accounts for healthy nodes becoming
infected: a randomly chosen node is healthy with probability
$1-\rho(t)$, this healthy node can be infected by either of its $k$
neighbor nodes with total probability of $\beta k \rho(t)$. The
stationary endemic state is reached when ${d \rho(t) / d t}= 0$
which leads to
\begin{equation}
\rho = 1 - \frac{1}{\beta k },
\label{SIregular2}
\end{equation}
indicating the existence of a nonzero epidemic threshold of
$\beta=1/k$. The innermost core of a network consisting only of nodes
with degrees $k \geq k_{S_{max}}$ will have epidemic threshold
\begin{equation}
\beta_c \leq 1 / k_{S_{max}}.
\end{equation}
The above inequality holds for all considered networks. Moreover, this
inequality becomes an equality for CNI and collaboration networks
where nearly all nodes in the innermost cores have degree $k \approx
k_{S_{max}}$.

\end{document}